\documentclass[useAMS,usenatbib,usegraphicx]{mn2e} 
\usepackage{amssymb} 
\usepackage{times} 
\usepackage{aas_macros} 
\usepackage[usenames]{color} 
\usepackage{appendix}
\usepackage{color}
\usepackage{float}
\usepackage{threeparttable}

\title[HD~86181]{Discovery of multiple p-mode pulsation frequencies in the roAp star, HD~86181 }

\author[Fangfei Shi et al.]{Fangfei Shi$^{1,2}$, Donald W. Kurtz$^{3,4}$, Daniel L. Holdsworth$^{4}$, Hideyuki Saio$^{5}$, \newauthor{Margarida~S.~Cunha$^{6}$, Huawei Zhang$^{1,2}$, Jianning Fu$^{7}$, G. Handler$^{8}$}
\\
$^{1}$Department of Astronomy, School of Physics, Peking University, Beijing 100871, P. R. China\\
$^{2}$Kavli Institute for Astronomy and Astrophysics, Peking University, Beijing 100871, P. R. China\\
$^{3}$Centre for Space Research, Physics Department, North-West University, Mahikeng 2745, South Africa\\
$^{4}$Jeremiah Horrocks Institute, University of Central Lancashire, Preston PR1 2HE, UK\\
$^{5}$Astronomical Institute, Graduate School of Science, Tohoku University, Sendai   980-8578, Japan \\
$^{6}$Instituto de Astrof\'\i sica e Ci\^encias do Espa\c co, Universidade do Porto, CAUP, Rua das Estrelas, PT4150-762 Porto, Portugal\\
$^{7}$Department of Astronomy, Beijing Normal University, Beijing 100875, P. R. China\\
$^{8}$Nicolaus Copernicus Astronomical Center, Polish Academy of Sciences, ul. Bartycka 18, 00-716, Warsaw, Poland\\
}

\date{Accepted XXX. Received YYY; in original form ZZZ}

\pubyear{2019}

\begin{document}
\label{firstpage}
\pagerange{\pageref{firstpage}--\pageref{lastpage}}
\maketitle

\begin{abstract} 
We report the frequency analysis of a known roAp star, HD~86181 (TIC~469246567), with new inferences from TESS data. We derive the rotation frequency to be $\nu_{\rm rot} = 0.48753 \pm 0.00001$\,d$^{-1}$. The pulsation frequency spectrum is rich, consisting of two doublets and one quintuplet, which we interpret to be oblique pulsation multiplets from consecutive, high-overtone dipole, quadrupole and dipole modes. The central frequency of the quintuplet is $232.7701$\,d$^{-1}$ (2.694 mHz). The phases of the sidelobes, the pulsation phase modulation, and a spherical harmonic decomposition all show that the quadrupole mode is distorted. Following the oblique pulsator model, we calculate the rotation inclination, $i$, and magnetic obliquity, $\beta$, of this star, which provide detailed information about the pulsation geometry. The $i$ and $\beta$ derived from the best fit of the pulsation amplitude and phase modulation to a theoretical model, including the magnetic field effect, slightly differ from those calculated for a pure quadrupole, indicating the contributions from $\ell=4, 6, 8, ...$ are small. Non-adiabatic models with different envelope convection conditions and physics configurations were considered for this star. It is shown that models with envelope convection almost fully suppressed can explain the excitation at the observed pulsation frequencies.
\end{abstract} 


\section{Introduction}
\label{intro}

The Ap (chemically peculiar A-type) stars have non-uniform distributions of chemical abundances on their surfaces and strong magnetic fields. These magnetic fields suppress surface convection that then leads to element stratification. For some heavy elements, such as Eu, Sr and Si, the radiation pressure can lift them up to the surface against gravity leading to many absorption features. These elemental overabundances occur in spots, making Ap stars obliquely rotating variable stars of a class known as $\alpha^2$~CVn stars \citep{1969ApJS...18..347P}.

Some cool Ap stars exhibit high-overtone, low-degree pressure pulsation modes with periods between 4.7 and 24\,min (frequencies in the range $55.8 - 300$\,d$^{-1}$; $0.6 - 3.5$\,mHz \citep{2021arXiv210513274H}) and photometric amplitudes up to 0.018 mag in Johnson $B$ \citep{2019MNRAS.487.3523C,2009CoAst.159...61K,2015MNRAS.452.3334S}. They are called rapidly oscillating Ap (roAp) stars. Some of these stars show both rotation features with periods of days to decades, and pulsation features in their light curves. 

\citet{1950MNRAS.110..395S} developed the oblique rotator model of the Ap stars, which accounts for the magnetic, spectral, and light variations observed in Ap stars. Following this model, \citet{1982MNRAS.200..807K} introduced the oblique pulsator model, which was generalized with the effects of both the magnetic field and rotation taken into account \citep{1982MNRAS.200..807K,1985ApJ...296L..27D,1993PASJ...45..617S,1994PASJ...46..301T,1995PASJ...47..219T,2004MNRAS.350..485S,2002A&A...391..235B,2011A&A...536A..73B}. According to this model, the pulsation axis is misaligned with the rotation axis, and generally closely aligned to the magnetic axis. When the star rotates, the viewing aspect of the pulsation modes varies along the line of sight, leading to apparent amplitude and phase modulation. This modulation can provide information on the geometry of observed pulsations, hence mode identification, which is necessary for asteroseismic inference with forward modelling.

Since the first roAp stars were discovered by \citet{1982MNRAS.200..807K}, 88 roAp stars have been found \citep{2015MNRAS.452.3334S, 2019MNRAS.488...18H,2019MNRAS.487.3523C,2019MNRAS.487.2117B,2021arXiv210513274H}. Asteroseismology is a useful method to diagnose stellar structure and interior physics from the evidence of surface pulsations \citep{2003MNRAS.343..831C}. Progress of this research for roAp stars has been hindered by the relatively small number of known stars, and because their rapid pulsation requires dedicated observations and high accuracy to detect the small pulsation amplitudes \citep{2019MNRAS.488...18H,2019MNRAS.487.3523C,2019MNRAS.487.2117B}. 

The space telescopes {\it Kepler} and the TESS (Transiting Exoplanet Survey Satellite) provide an opportunity to detect oscillations well below the amplitude threshold of ground-based observations. Both {\it Kepler} and TESS have short cadence (2 min for TESS and 58.89\,s for {\it Kepler})  observations, but {\it Kepler} only observed 512 stars in this mode during each observing `quarter'. However, the standard long cadence sampling frequency of the {\it Kepler} 30-min observations is generally too low for studying the pulsation of roAp stars in detail. 

\citet{2013MNRAS.430.2986M} showed that the Nyquist ambiguity in the LC data can be resolved as a result of the Barycentric corrections applied to {\it Kepler} time stamps, and \citet{2019MNRAS.488...18H} discovered 6 roAp candidates through this method. Compared to the  {\it Kepler} 58.89-s observations, TESS is observing many more stars with 2-min observations with sufficiently long time bases to detect pulsations. Up to now, 21 new roAp stars have been found from just TESS sectors 1 to 13 \citep{2019MNRAS.487.3523C,2019MNRAS.487.2117B,2021arXiv210513274H}.

Before the TESS observations of our target, HD~86181, \citet{1994IBVS.4013....1K} discovered it to be a roAp star from 4.85\,hr of ground-based data. They reported the star to have a pulsation period of 6.2\,min and with an amplitude of 0.35\,mmag through a Johnson $B$ filter. That period corresponds to a frequency of 2.688\,mHz, or 232.26\,d$^{-1}$. No further detailed studies of the pulsations in HD~86181 have been published.

Parameters for this star are listed in Table~\ref{Tab:param}. The effective temperature was estimated using the Str$\ddot{o}$mgren photometric indices extracted from the catalogue of \citet{1998AandAS..129..431H} and the calibrations in the TEMPLOGG code (Rogers\,1995) which were developed based on the work of \citet{1985MNRAS.217..305M} and \citet{1993AandA...268..653N}. Since no convincing uncertainty is given by this method, we indicate, instead, a range of values of $T_{\rm eff}$ published in the literatures. 

The luminosity was calculated through the relation $-2.5\log L=M_G+BC_G(\rm T_{eff})-M_{bol,\odot}$, where $BC_G(\rm T_{eff})$ is a temperature dependent bolometric correction defined in \citet{2018AandA...616A...8A}, and the uncertainty of BC (Bolometric Correction) is 0.13, based on a comparison with Ap data that is described in some detail in \citet{2019MNRAS.487.3523C}. 
While the uncertainty derived in \citet{2019MNRAS.487.3523C} was based on a comparison of Ap-star measurements with the empirical BC$_V$ calibration by \citet{1996ApJ...469..355F} and, thus, the consistency of using it with the BG$_G$ values derived from  the calibration of  \citet{2018AandA...616A...8A} may be questionable, it provides a more conservative result than the uncertainty derived from \citet{2018AandA...616A...8A}, which does not account for the stars' peculiarities.
The extinction in the $G$ band used to calculate $M_G$ here was from \citet{2019AandA...628A..94A}, and the uncertainty is 0.2, which is the value indicated in the Figure 20 in \citep{2019AandA...628A..94A}. The parallax was from Gaia eDR3 \citep{2020yCat.1350....0G}. $M_{\rm bol,\odot}$ adopted is 4.74 as defined by IAU Resolution 2015 B2\footnote{https://www.iau.org/static/resolutions/IAU2015\_English.pdf}.

\begin{table}

\scriptsize
\centering
\caption{Parameters of HD~86181.} 
\begin{tabular}{lll}
\hline
\hline
Apparent $G$ magnitude & $9.341\pm0.003$ &  \citet{2020yCat.1350....0G}\\
Extinction in $G$ band & $0.1\pm0.2$ &  \citet{2019AandA...628A..94A}\\
Spectral type & F0 Sr &\citet{2009AandA...498..961R} \\
Parallax (mas) & $4.15 \pm 0.01$ & \citet{2020arXiv201201533G}\\
Distance (pc) & $241.0 \pm 0.6$ & derived from parallax \\
 $b-y$ & 0.175 & \citet{1991PASP..103..494P}\\
$m_1$ &  0.245 & \citet{1991PASP..103..494P}\\
$c_1$ & 0.702 & \citet{1991PASP..103..494P}\\
H$_\beta$ & 2.804 & \citet{1991PASP..103..494P}\\
$T_{\rm eff}$(K) & $7750$; [7240-7910] & This work$^*$; Literature$^+$\\
Luminosity (L$_\odot$) & $8.8 \pm 1.9$ & \citet{2018AandA...616A...8A}\\
Mean longitudinal  & $536 \pm 75$ & \citet{2015AandA...583A.115B} \\
magnetic field (G) & & \\
\hline
\hline
\end{tabular}
\label{Tab:param}
 \begin{tablenotes}
        \footnotesize
        \item $^*$ based on Rogers 1995
        \item $^+$ \citet{2020AandA...636A..74T}, \citet{2019AandA...628A..94A}, \citet{1996AandA...311..901M}
      \end{tablenotes}
\end{table}

\section{TESS observations}
\label{obs}

HD~86181 was observed by TESS in sectors 9 and 10 in 2-min cadence. The data have a time span of 51.76\,d with a centre point in time of $t_0 = {\rm BJD}~2458569.80077$, and comprise 33832 data points after some outliers were removed. The standard PDC SAP (pre-search data conditioning simple aperture photometry) fluxes provided by MAST (Mikulski Archive for Space Telescopes) were used and normalised by dividing by the median flux separately for each sector. Relative magnitudes were then calculated from the processed fluxes, giving the light curve shown in the top panel of Fig.\,\ref{fig:lc1}.

There are obvious rotational variations from spots, as is typical of the magnetic Ap stars. Within the oblique rotator model, the double wave nature of the rotational variations suggests that two principal spots with enhanced brightness on the stellar surface are seen. The high frequency pulsation cannot be seen in this figure at this resolution.

\section{Frequency analysis}
\subsection{Rotation frequency analysis}
\label{fr}

Before we conducted a detailed analysis of the rotation frequency of HD\,86181, we first measured the rotation frequency with a coarse Discrete Fourier Transform \citep[DFT;][]{1985MNRAS.213..773K} such that we could bin the data every one rotation cycle. This allowed us to assess the instrumental variation, which we subsequently fit with a polynomial and removed from the original light curve. We then calculated a DFT with a finer frequency grid, as shown in Fig.\,\ref{fig:ft1}, to measure the stellar rotation frequency. 
The low frequencies dominate in the spectrum, so we zoom in to both the low frequency range (second panel) and high frequency range (third panel). From the amplitude spectrum at low frequency, the rotational harmonics are clearly seen. Although the highest peak is at a frequency of 0.97\,d$^{-1}$, considering the phase plot, we derive the rotation frequency to be around 0.48\,d$^{-1}$. Because the variation is a double wave, the second harmonic has the highest amplitude.

A linear least-squares fit was calculated to find the best amplitudes and phases of the rotation frequency and its 4 visible harmonics, and then a non-linear least-squares fit to get optimized results. The rotational frequency is derived to be $\nu_{\rm rot} = 0.48753 \pm 0.00001$\,d$^{-1}$ ($P_{\rm rot} = 2.05116 \pm 0.00004$\,d) by dividing the frequency of the highest amplitude second harmonic by two, which has better signal-to-noise ratio. Besides the rotation frequency and its harmonics, there are still some signals left in the low frequency range, probably because instrumental variation has not been removed completely. These signals were removed prior to the non-linear least square fits for better estimates of the uncertainties. The uncertainties were derived following \citet{1999DSSN...13...28M}. The rotation period is short among the known roAp stars, after HD~43226 (\citealt{2019MNRAS.487.3523C}), HD~216641 (\citealt{2019MNRAS.487.3523C}), and HD~6532 (\citealt{1996MNRAS.280....1K}, \citealt{1996MNRAS.281..883K}), which have similar rotation periods of $P_{\rm rot} = 1.71441$\,d, $P_{\rm rot} = 1.876660$\,d, and $P_{\rm rot} = 1.944973$\,d, respectively.

\begin{figure}
\centering
\includegraphics[width=0.95\linewidth,angle=0]{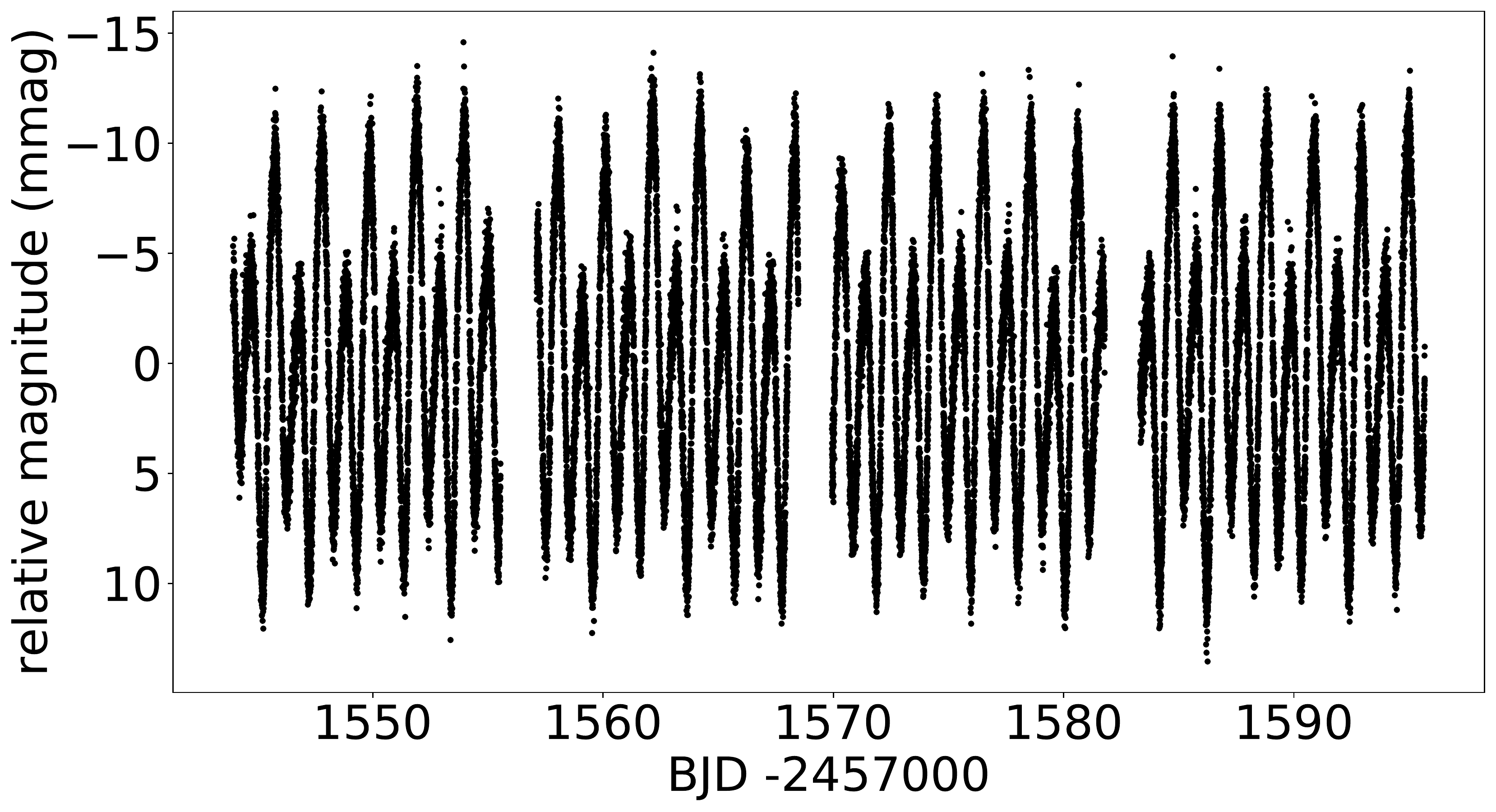}	
\includegraphics[width=0.95\linewidth,angle=0]{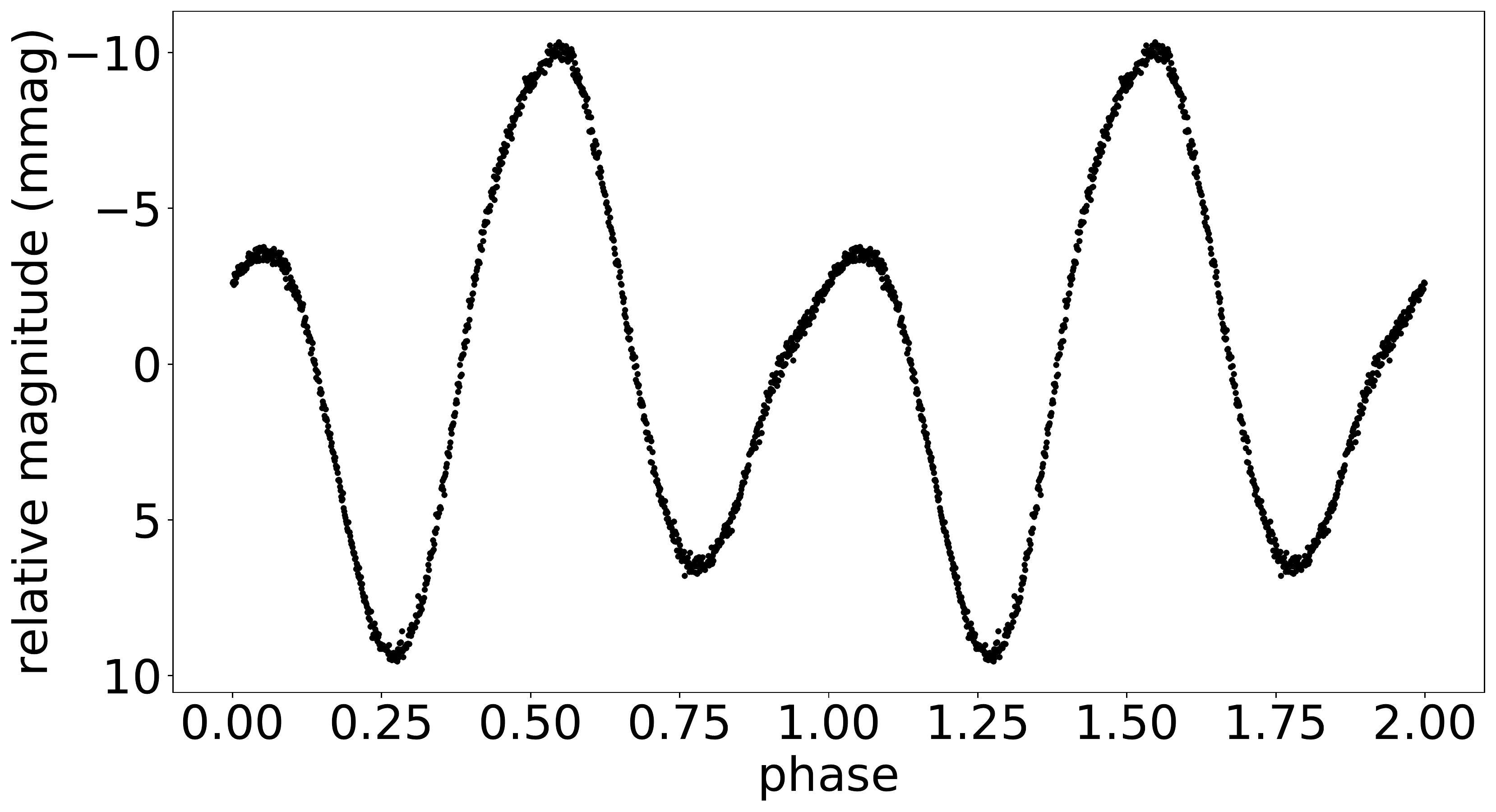}	
\caption{Top: The light curve of HD~86181 showing the rotational variations. Bottom: Phase folded light curve of HD~86181, folded on the rotation period of 2.05116\,d; two rotation cycles are shown for clarity. The data are from TESS sectors 9 and 10. The time zero-point, BJD 2458569.26128, is the time of pulsation maximum. The phases are binned every 0.001 phase bin. }
\label{fig:lc1}
\end{figure}

\begin{figure}
\centering
\includegraphics[width=0.80\linewidth]{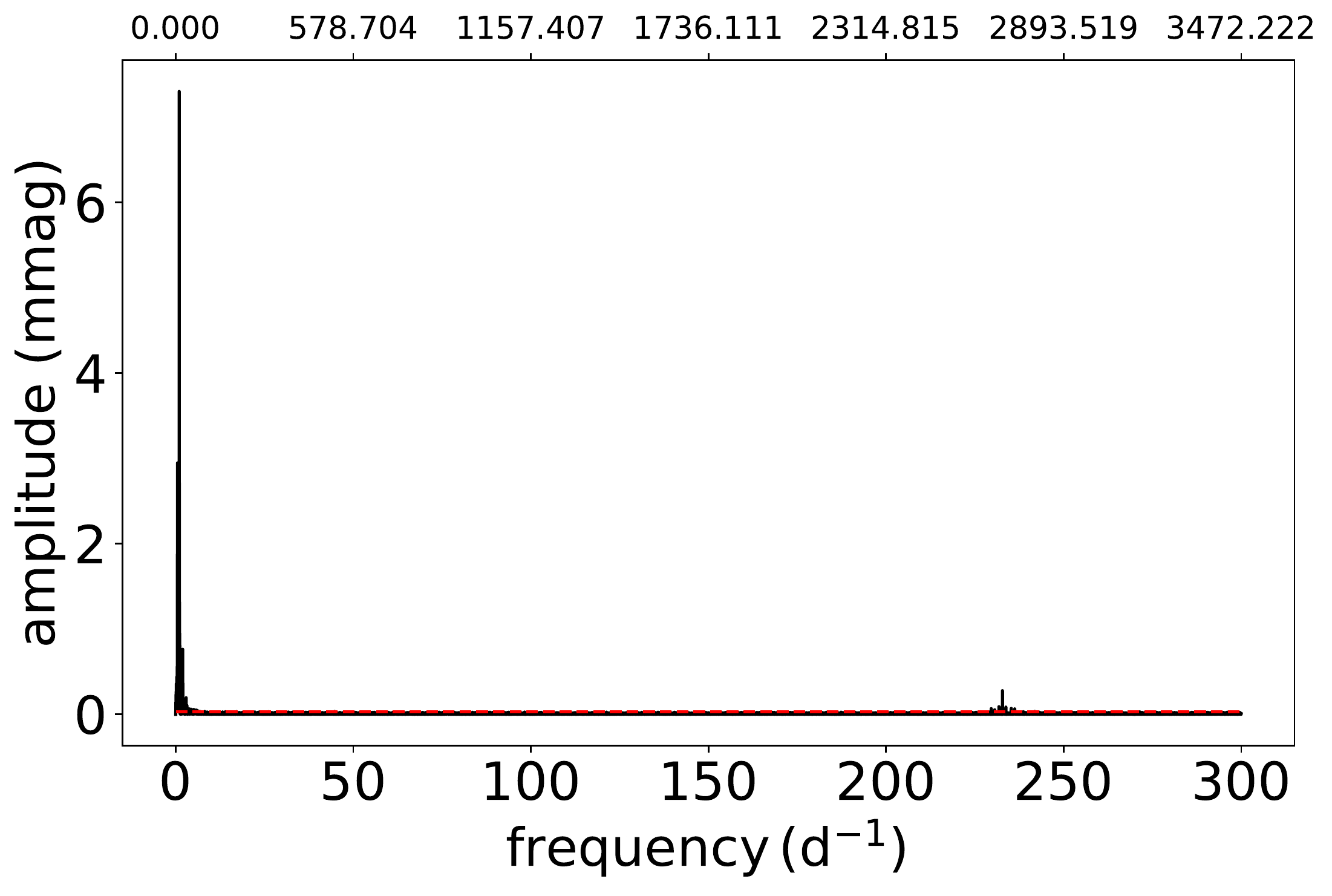}	
\includegraphics[width=0.82\linewidth]{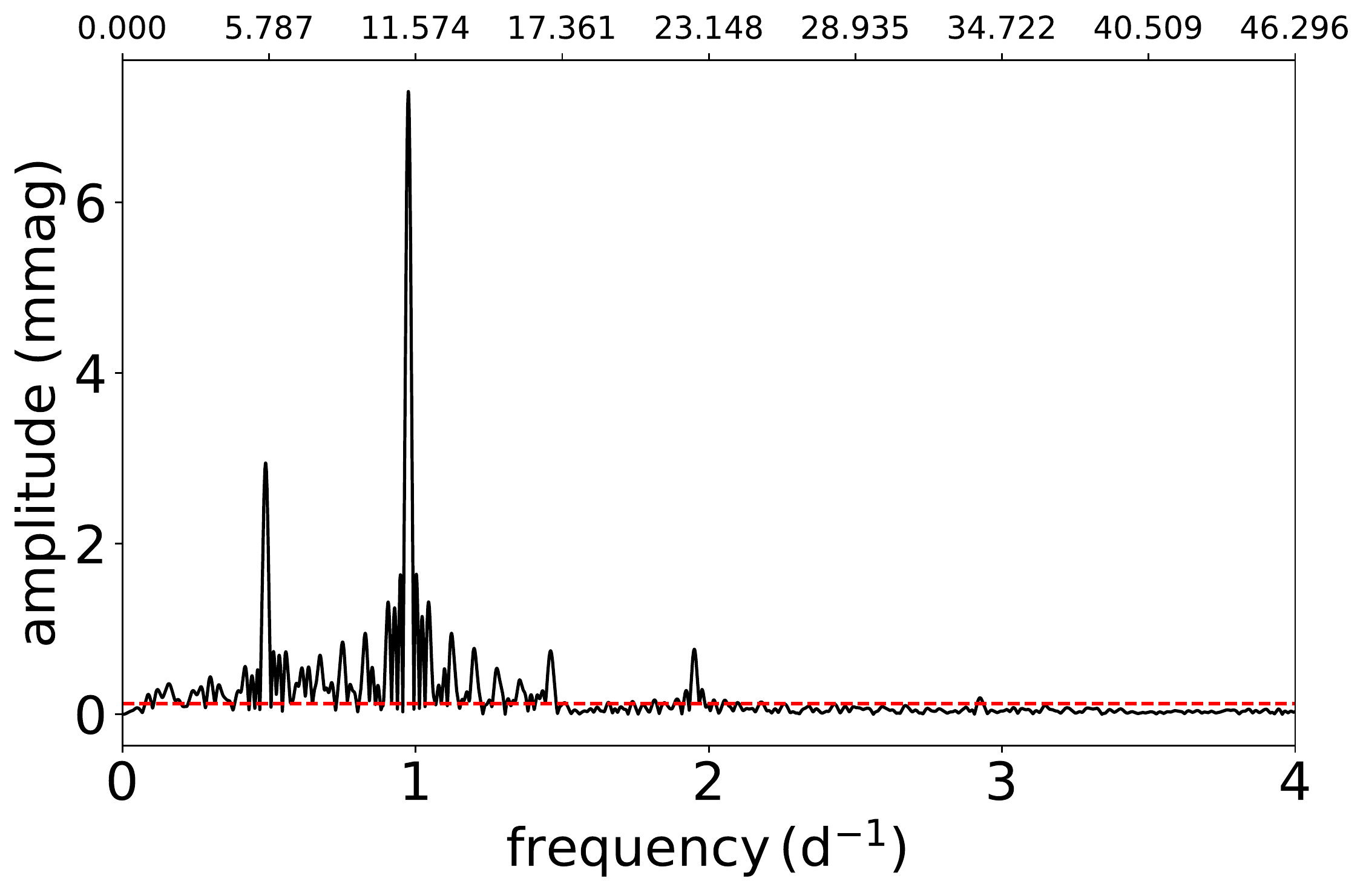}	
\includegraphics[width=0.80\linewidth]{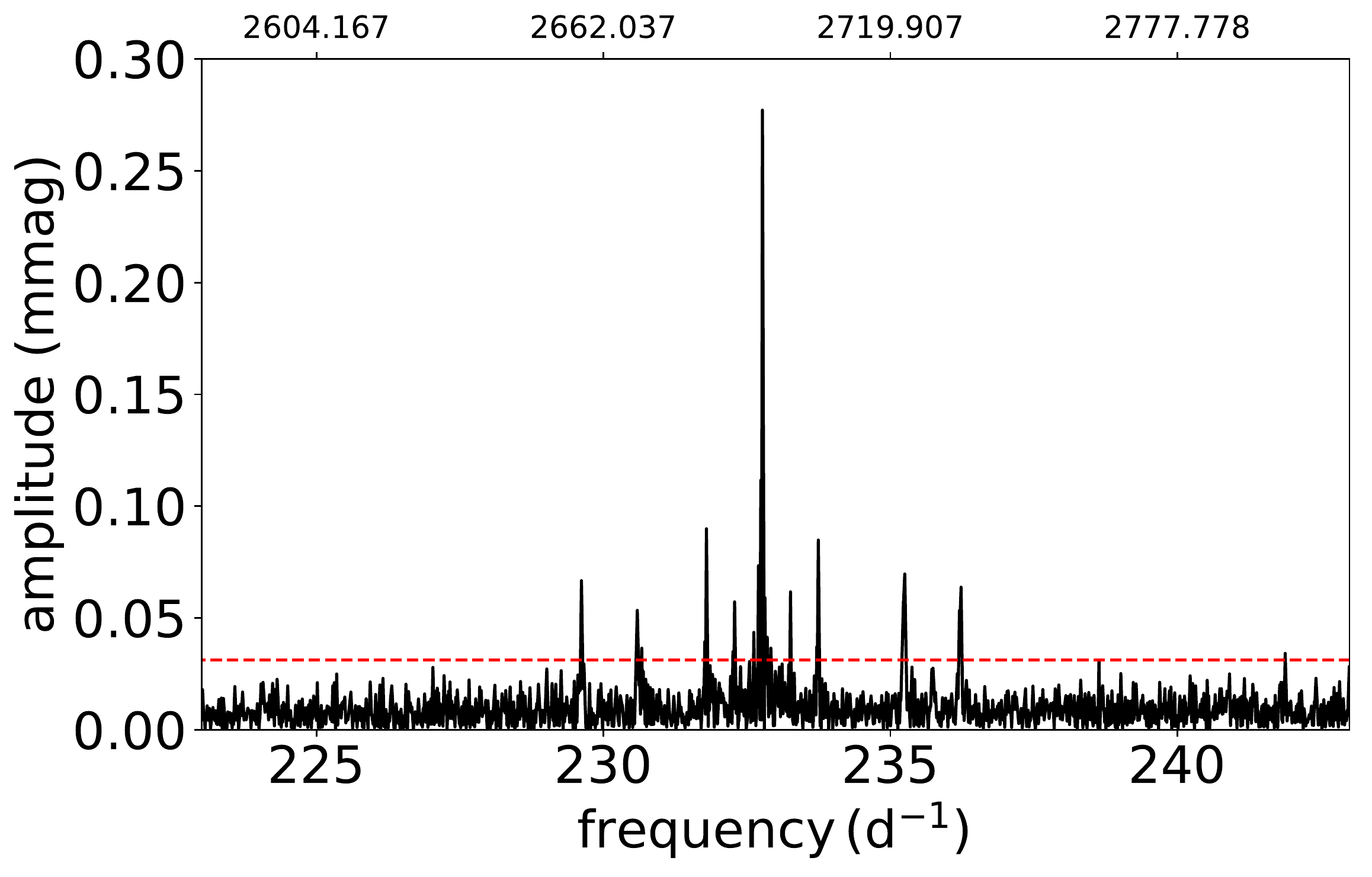}	
\includegraphics[width=0.80\linewidth]{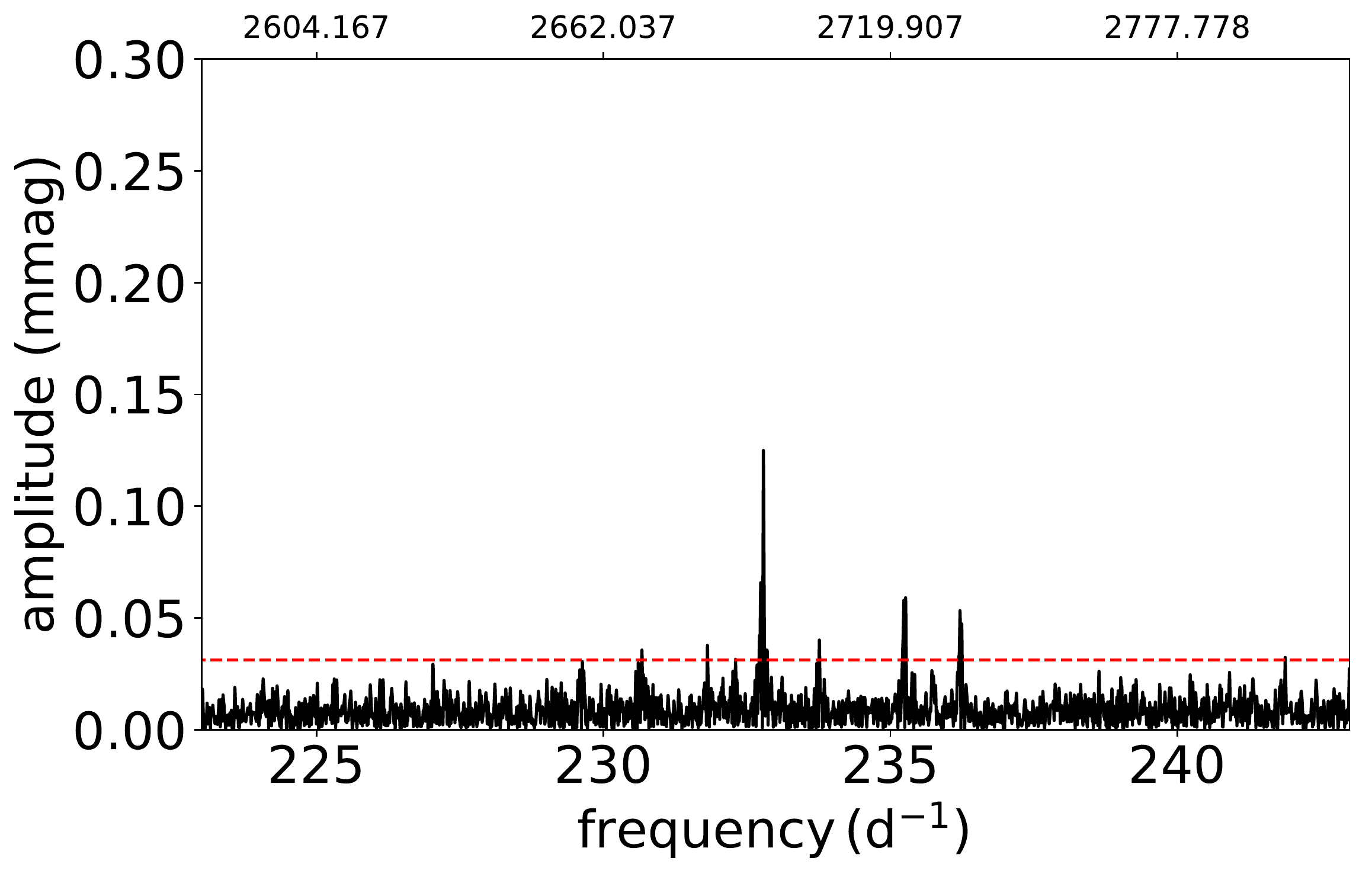}	
\caption{The frequency spectrum of HD~86181. Top: The amplitude spectrum of the S9--10 data out to 300\,d$^{-1}$. The rotational frequencies at low frequency dominate. The pulsation frequencies centred on 232.2\,d$^{-1}$ are difficult to see at this scale. Second: The low frequency rotational harmonics. Third: the pulsation frequencies for the high-pass filtered data. Bottom: The frequency spectrum after the frequencies in Table~2 have been removed. The red horizontal lines are 4 times of noise level. The top x-axis is the corresponding frequency in $\umu$Hz.}
\label{fig:ft1}
\end{figure}

\subsection{The pulsations}
\label{fp}

To study the pulsations, a high-pass filter was used to remove the rotational light variations, any remaining instrumental artefacts and other low frequencies. The high-pass filter was a simple consecutive pre-whitening of low frequency peaks extracted by Fourier analysis until the noise level was reached in the frequency range $0 - 6$\,d$^{-1}$. The third panel in Fig.\,\ref{fig:ft1} shows the amplitude spectrum for the high-pass filtered data around the high-frequency variability. By inspection it can be seen that there is a central quintuplet and two doublets, one at higher and another at lower frequency than the quintuplet. After removing these three groups of frequencies, five singlets still remain (see the bottom panel of Fig.\,\ref{fig:ft1}). However, their frequencies are similar to the quintuplet and two doublets within the uncertainties. These may be caused by amplitude or frequency modulation over the time span of the data set, 51.76\,d. 

\begin{figure}
\centering
\includegraphics[width=1.0\linewidth]{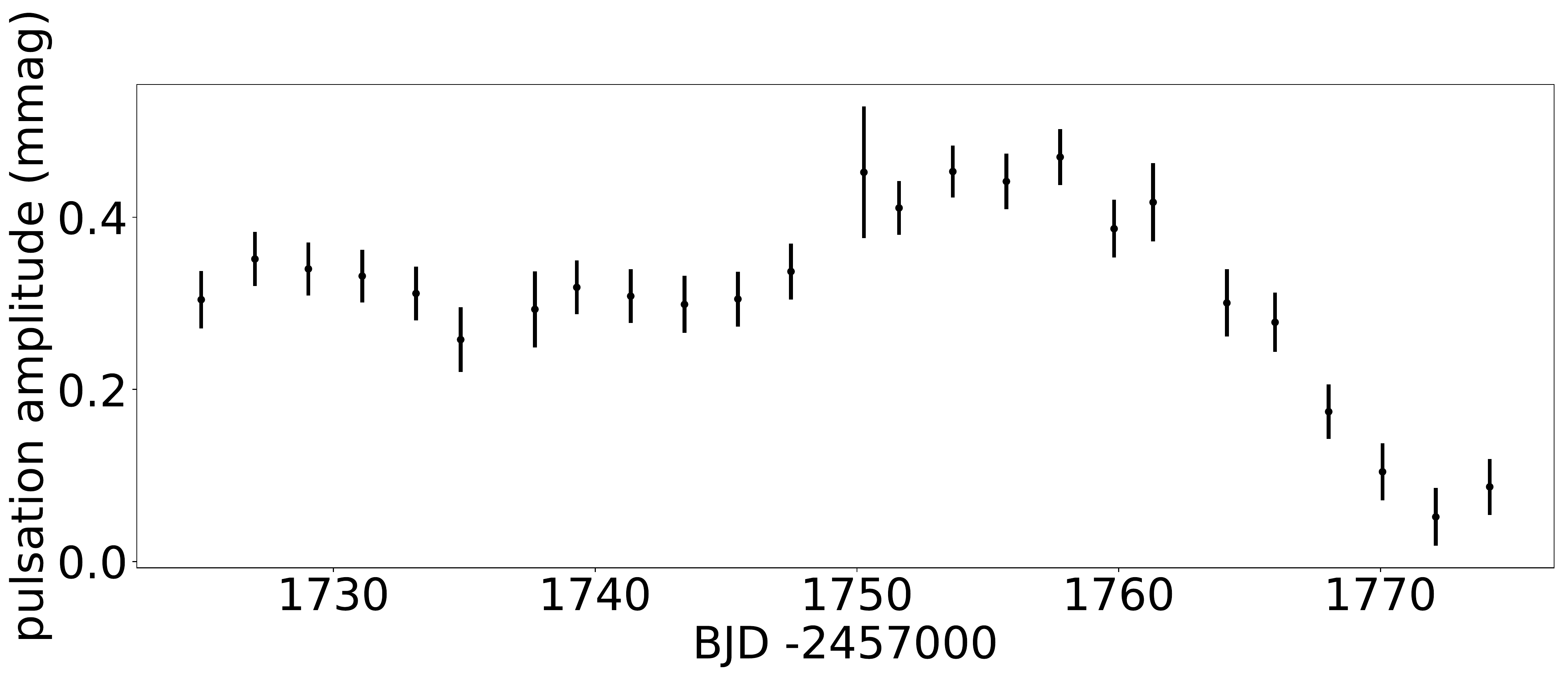}	
\includegraphics[width=1.0\linewidth]{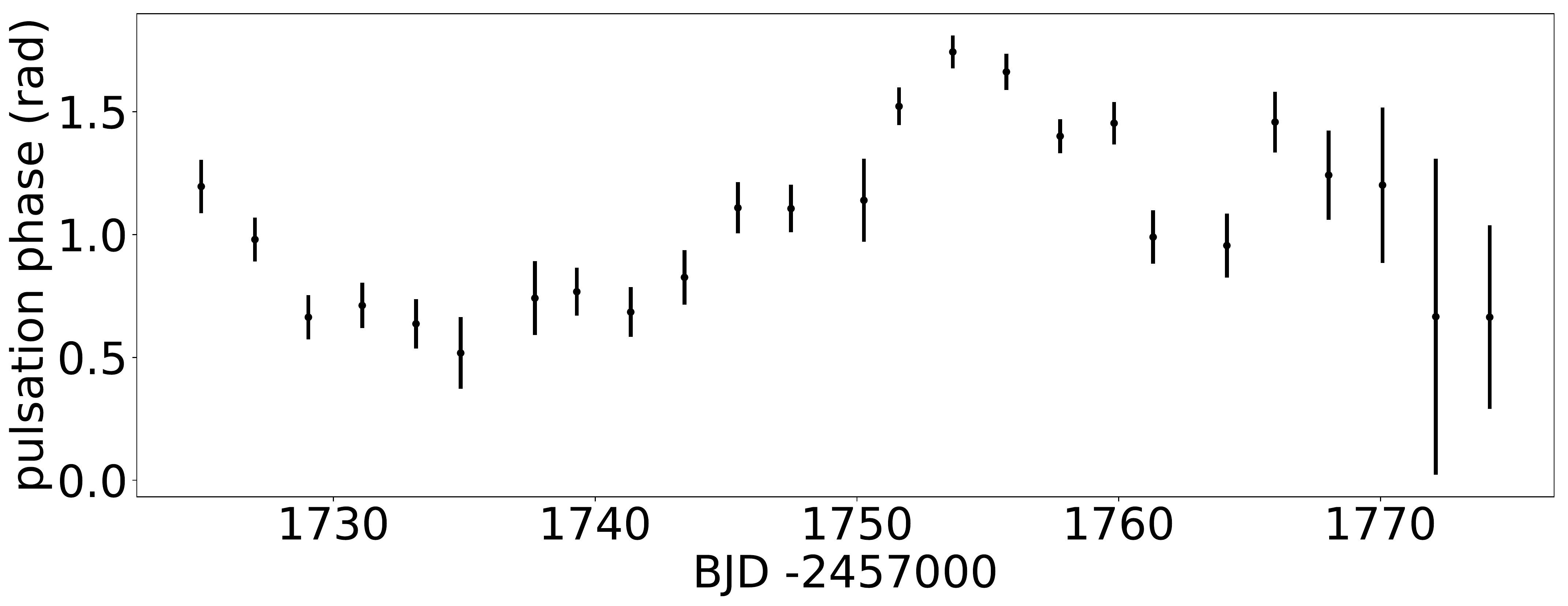}	
\caption{The pulsation amplitude and phase variations of HD~86181 for the dominant quadrupole mode. Top: pulsation amplitude variations as a function of time.  Bottom: pulsation phase variations as a function of time. }
\label{fig:modulation_1rot}
\end{figure}

To test this, we removed the doublets and singlets from the light curve and fitted $\nu_1$ to sections of the data that are exactly one rotation cycle long and calculated the amplitude and phase. Fig.\,\ref{fig:modulation_1rot} shows there is amplitude and phase variability with time. By choosing exactly one rotation length of data, the amplitude and phase variations due to oblique pulsation were smoothed.

If the frequency were stable, there would be no phase variations. As the data were fitted with the function $\Delta m = A\cos(\nu(t-t_0)+\phi)$, the frequency and phase terms are inextricably intertwined \citep[see the section 5.3.2 in][]{2014MNRAS.443.2049H}, thus a change in one can be interpreted as a change in the other. Therefore, although we show a change in the phase in Fig. 3, the change could be in the frequency.  Such variability is common in roAp stars studied with high precision data \citep{2021FrASS...8...31H}.

As in the analysis of rotation frequency, linear and non-linear least squares fits were used to get optimised results of frequencies, amplitudes and phases. The non-linear least squares fit results are shown in Table~\ref{Tab:nld}. Within the uncertainties, the sidelobes of the quintuplet are exactly split by the rotation frequency. In addition to the quintuplet, there are two doublets that are split by 2$\nu_{\rm rot}$; these are the sidelobes of two dipole pulsation frequencies that are labeled as $\nu_2$ and $\nu_3$. For a pure dipole or quadrupole pulsation, the oblique pulsator model requires that the sidelobes are split by exactly the rotation frequency of the star, and that the phases of all components are equal at the time of pulsation maximum. To test this, the frequency of the quintuplet sidelobes were fixed to be equally spaced by the rotation frequency, and the zero-point in time was chosen such that the phases of the first pair of sidelobes are the same, then a linear least squares fit was applied to the data with the results show in Table~\ref{Tab:ls}. The phases of the quintuplet sidelobes are not equal within the uncertainties, which indicates this star pulsates in a distorted quadrupole mode.

\begin{table}
\centering
\caption{A non-linear least squares fit of the frequency multiplets for HD~86181. The zero point for the phases is $t_0 = {\rm BJD}~2458569.26128$.}
\begin{tabular}{rrcr}
\hline
&\multicolumn{1}{c}{frequency} & \multicolumn{1}{c}{amplitude} &   
\multicolumn{1}{c}{phase}  \\
&\multicolumn{1}{c}{d$^{-1}$} & \multicolumn{1}{c}{mmag} &   
\multicolumn{1}{c}{radians}   \\
& & \multicolumn{1}{c}{$\pm 0.007$} &   
   \\
\hline
$\nu_{rot}$ &  $0.48765 \pm 0.00003$  &  2.970  &  $5.829  \pm  0.003$ \\ 
$2\nu_{rot}$ & $0.97506 \pm 0.00001$  &  7.296  &  $2.776  \pm  0.001$ \\ 
$3\nu_{rot}$ &  $1.46233 \pm 0.00013$  &  0.732  &  $0.312  \pm  0.013$ \\ 
$4\nu_{rot}$ &  $1.95043 \pm 0.00013$  &  0.761  &  $6.177  \pm  0.013$ \\ 
$6\nu_{rot}$ &  $2.92585 \pm 0.00050$  &  0.190  &  $0.215  \pm  0.049$ \\ 
\hline
$\nu_2 - \nu_{\rm rot}$  &  $229.6162 \pm 0.0012$  &  0.059  &  $0.28  \pm  0.17$ \\ 
$\nu_2 + \nu_{\rm rot}$ &  $230.5897 \pm 0.0014$  &  0.050  &  $0.49  \pm  0.20$ \\ 
\hline
$\nu_1 - 2\nu_{\rm rot}$
&  $231.7947 \pm 0.0008$  &  0.091  &  $6.00  \pm  0.11$ \\ 
$\nu_1 - \nu_{\rm rot}$
&  $232.2853 \pm 0.0013$  &  0.055  &  $6.27  \pm  0.18$ \\ 
$\nu_1$ &  $232.7701 \pm 0.0003$  &  0.273  &  $6.24  \pm  0.04$ \\ 
 $\nu_1 + \nu_{\rm rot}$
&  $233.2587 \pm 0.0011$  &  0.062  &  $6.17  \pm  0.16$ \\ 
 $\nu_1 + 2\nu_{\rm rot}$
&  $233.7438 \pm 0.0008$  &  0.080  &  $0.14  \pm  0.12$ \\ 
\hline
$\nu_3 - \nu_{\rm rot}$
 &  $235.2495 \pm 0.0010$  &  0.071  &  $6.11  \pm  0.14$ \\ 
$\nu_3 + \nu_{\rm rot}$
 &  $236.2261 \pm 0.0012$ &  0.063  &  $5.94  \pm  0.16$ \\ 
 \hline
\hline
\end{tabular}
\label{Tab:nld}
\end{table}

\begin{table}
\centering
\caption{A least squares fit of the frequency multiplets for HD~86181, where the frequency splitting of the rotational sidelobes has been forced to be exactly the rotation frequency. The zero point for the phases, $t_0 = {\rm BJD}~2458569.26128$, has been chosen to be a time when the first two orbital sidelobes of the quintuplet have equal phase.} 
\begin{tabular}{rrcr}
\hline
&\multicolumn{1}{c}{frequency} & \multicolumn{1}{c}{amplitude} &   
\multicolumn{1}{c}{phase}  \\
&\multicolumn{1}{c}{d$^{-1}$} & \multicolumn{1}{c}{mmag} &   
\multicolumn{1}{c}{radians}   \\
& & \multicolumn{1}{c}{$\pm 0.007$} &   
   \\
\hline
$\nu_2 - \nu_{\rm rot}$  &  229.6162  &  0.058  &  $0.25  \pm  0.11$ \\ 
$\nu_2 + \nu_{\rm rot}$ &  230.5913  &  0.049  &  $0.40  \pm  0.14$ \\ 
\hline
$\nu_1 - 2\nu_{\rm rot}$
&  231.7950  &  0.091  &  $-0.26  \pm  0.07$ \\ 
$\nu_1 - \nu_{\rm rot}$
&  232.2826  &  0.053  &  $-0.13  \pm  0.12$ \\ 
$\nu_1$ &  232.7701  &  0.273  &  $-0.05  \pm  0.02$ \\ 
 $\nu_1 + \nu_{\rm rot}$
&  233.2576  &  0.061  &  $-0.13  \pm  0.11$ \\ 
 $\nu_1 + 2\nu_{\rm rot}$
&  233.7452  &  0.080  &  $0.14  \pm  0.08$ \\ 
\hline
$\nu_3 - \nu_{\rm rot}$
 &  235.2494  &  0.071  &  $-0.13  \pm  0.09$ \\ 
$\nu_3 + \nu_{\rm rot}$
 &  236.2245  &  0.063  &  $-0.11  \pm  0.11$ \\ 
 \hline
\hline
\end{tabular}
\label{Tab:ls}
\end{table}

We also investigated the impact of the spots on the pulsations. From the second panel of Fig.\,\ref{fig:lc1}, the rotational variation caused by the spots amounts to 20\,ppt peak-to-peak. We therefore expect the modulation of the pulsation caused by spots to be also a factor of 0.02 of the pulsation amplitude, which is down to $\mu$mag, much below the noise level. So the effect of spots on the pulsation amplitude is negligible.

Finally, harmonics of the pulsation frequencies were also searched for beyond the Nyquist frequency, $\nu_{Ny} = 359.804$\,d$^{-1}$. Only three similar alias groups centred at $2 \nu_{Ny}-\nu_1$, $2 \nu_{Ny}-\nu_2$ and $2 \nu_{Ny}-\nu_3$ were found, with no evidence of harmonics of pulsation frequencies.

\subsection{Pulsation amplitude and phase modulation }
\label{modulation}

\begin{figure}
\centering
\includegraphics[width=1.0\linewidth]{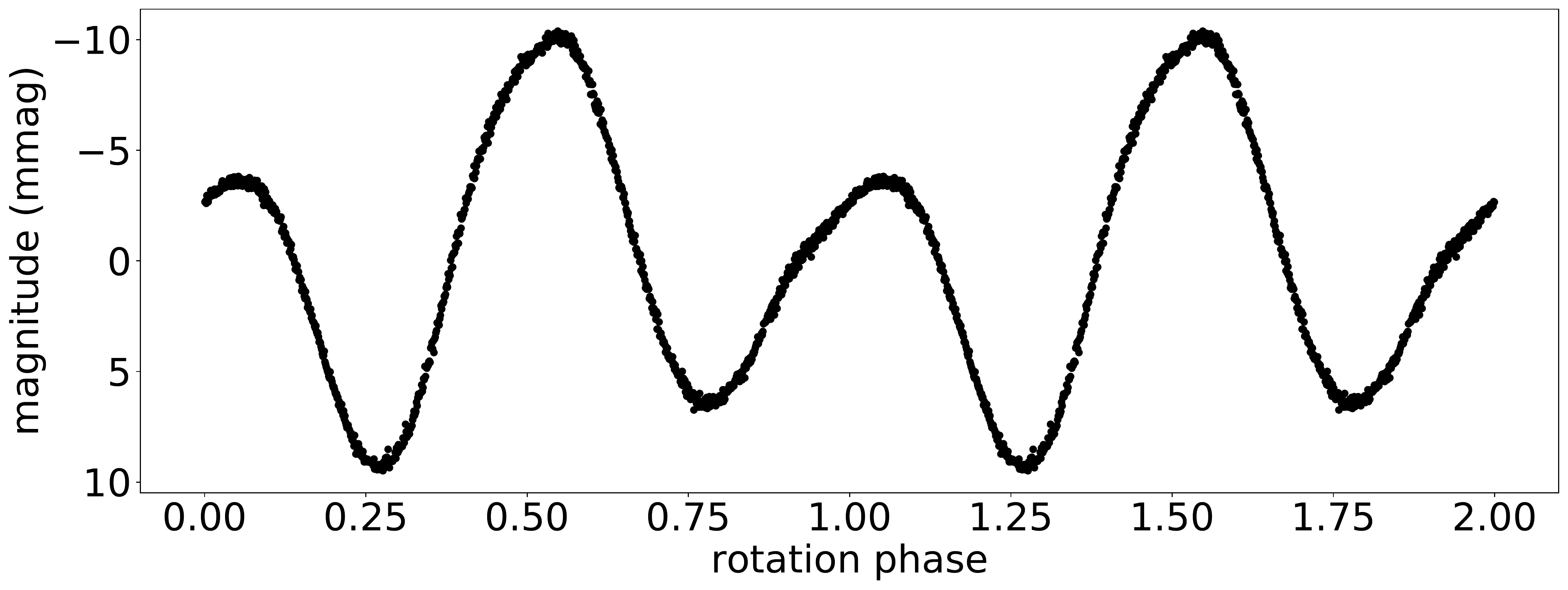}	
\includegraphics[width=1.0\linewidth]{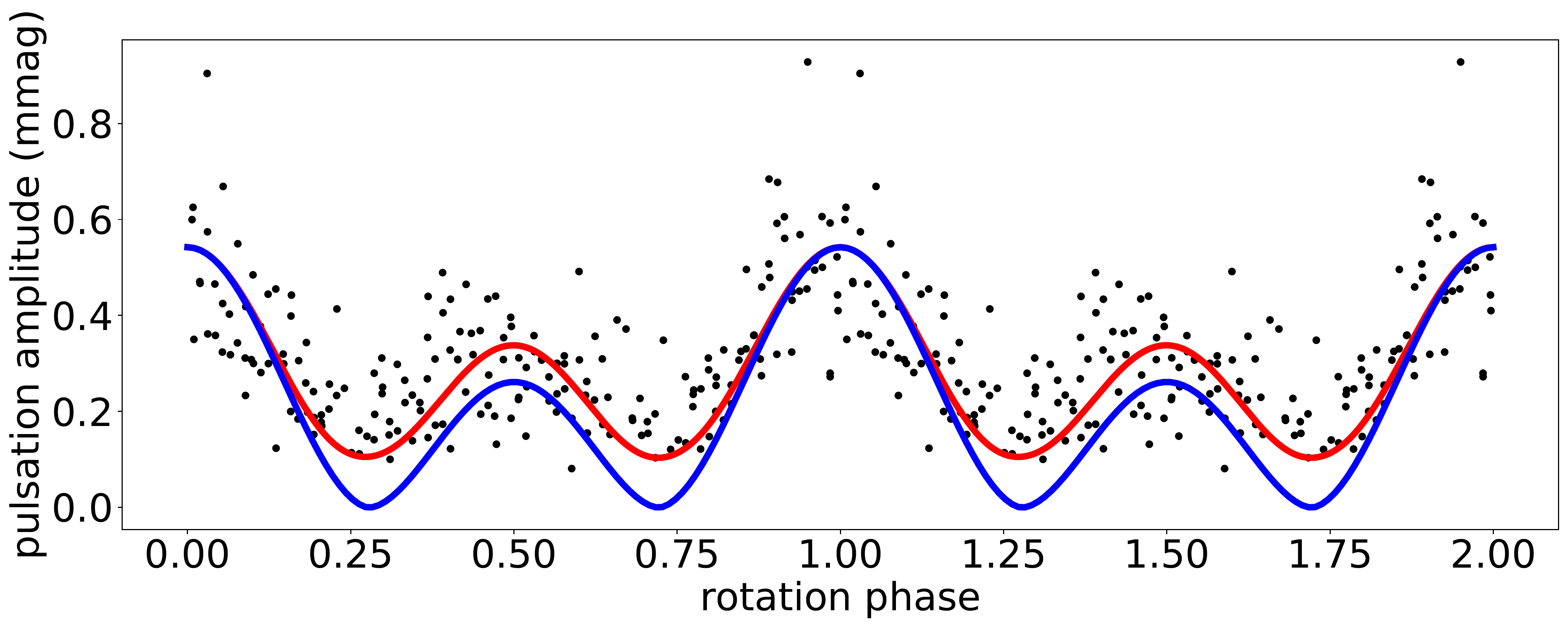}	
\includegraphics[width=1.0\linewidth]{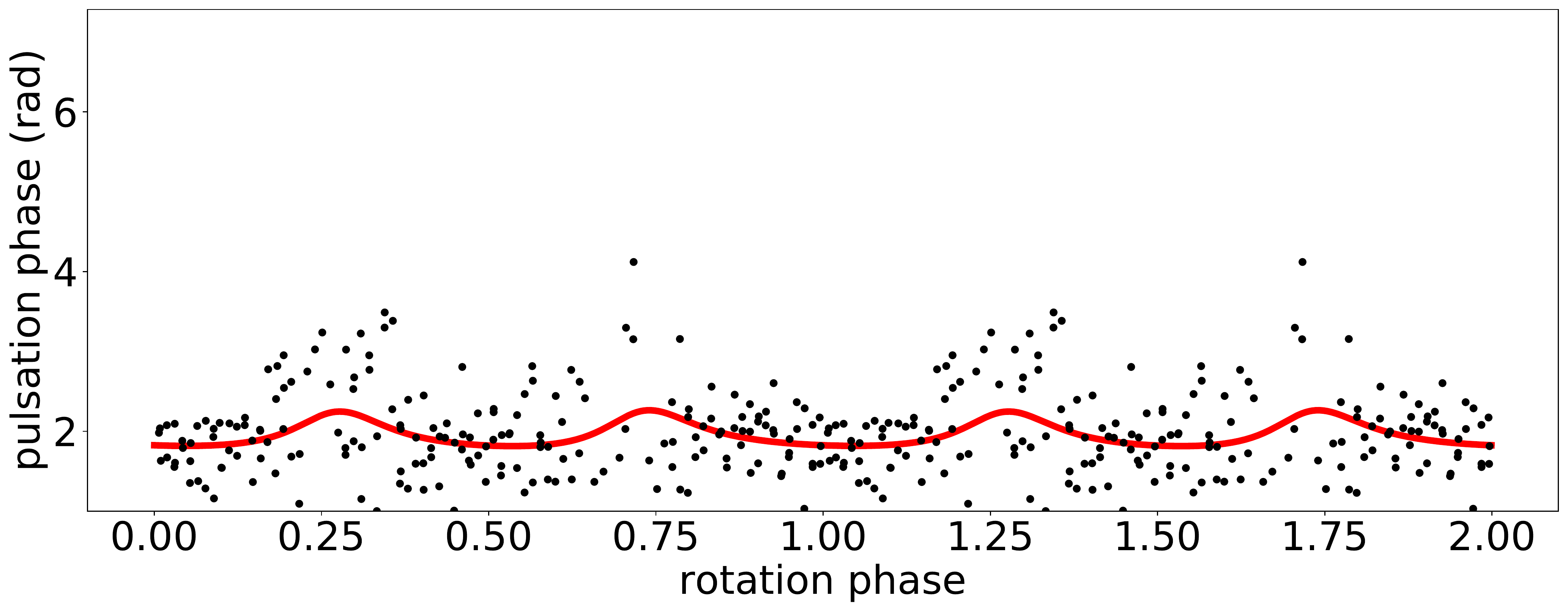}	
\caption{The pulsation modulation for pulsation frequency $\nu_1$ of HD~86181. Top: The phase folded rotation light curve. Middle:  pulsation amplitude variations as a function of rotation phase. Amplitude points with {$1\sigma$} errors greater than 0.12\,mmag are not plotted here. Bottom: pulsation phase variations as a function of rotation phase. Phase points with {$1\sigma$} errors greater than 0.8 rad are not plotted here. The red lines are theoretical amplitude modulation modelled following \citet{1992MNRAS.259..701K} with the components from Table~4. The blue line was calculated based on an oblique pure quadruple mode (see section~\ref{ib_cal}). Two rotation cycles are shown. The time zero-point is $t_0 = {\rm BJD}~2458569.26128$. }
\label{fig:modulation}
\end{figure}

To study the rotation modulation of the pulsation amplitudes and phases, the light curve was divided into 217 segments each containing 50 pulsation cycles, thus each segment had a time span of 0.21d, or 0.1 of a rotation cycle. Linear least-squares fitting was applied to these segments at fixed frequency, $\nu_1 = 232.7701$\,d$^{-1}$, to calculate the pulsation amplitude and phase as a function of rotation phase. Fig.\,\ref{fig:modulation} shows these modulations along with the rotation light variations for comparison.

The maxima of the pulsation amplitude depend on the aspect of the pulsation axis, while the light extrema depend on the spots. The difference between the occurrences of the extrema of the pulsation amplitude and the rotational light variations indicates the position of spots relative to the pulsation axis. In many Ap stars the surface positions where spots form -- particularly for the rare earth elements -- is related to the magnetic field. In the case that the spots are centred on the pulsation axis which is also fixed close to the magnetic axis, the rotation phase of pulsation maximum coincides with, or is near to, the rotation phase of the light extrema. As \citet{Handler_2006} showed for HD~99563, the maximum of pulsation amplitude coincides with the maximum of rotation light in red filters, and the minimum in blue filters. The antiphase variations in blue and red filters are related to the flux redistribution from UV to optical caused by line blocking \citep{1973ApJ...185..577L}.


For HD~86181, pulsation amplitude maximum coincides with the secondary maximum of the light curve, and after half a cycle, the secondary maximum of pulsation amplitude coincides with the maximum of the light curve. For a pure quadrupole pulsator, the intrinsic pulsation amplitude peaks at both pulsation poles and at the equator. The pulsation maximum at the poles is twice that at the equator, but with inverse phase. We assume the maximum of pulsation amplitude is generated at the pole, while the secondary maximum by equator. This assumption is verified with the the oblique pulsator model below.

\begin{figure}
\centering
\includegraphics[width=1.0\linewidth]{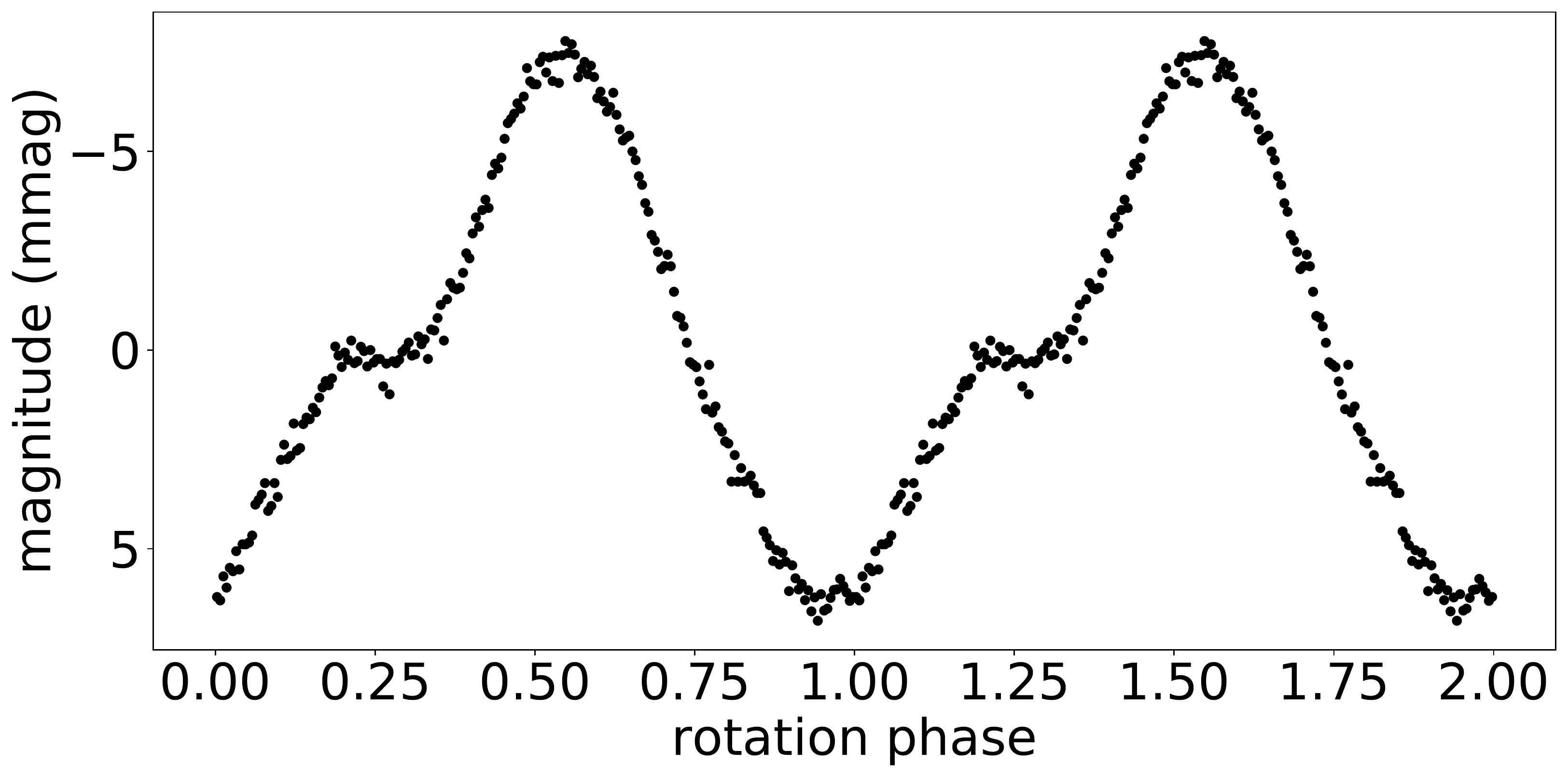}	
\includegraphics[width=1.0\linewidth]{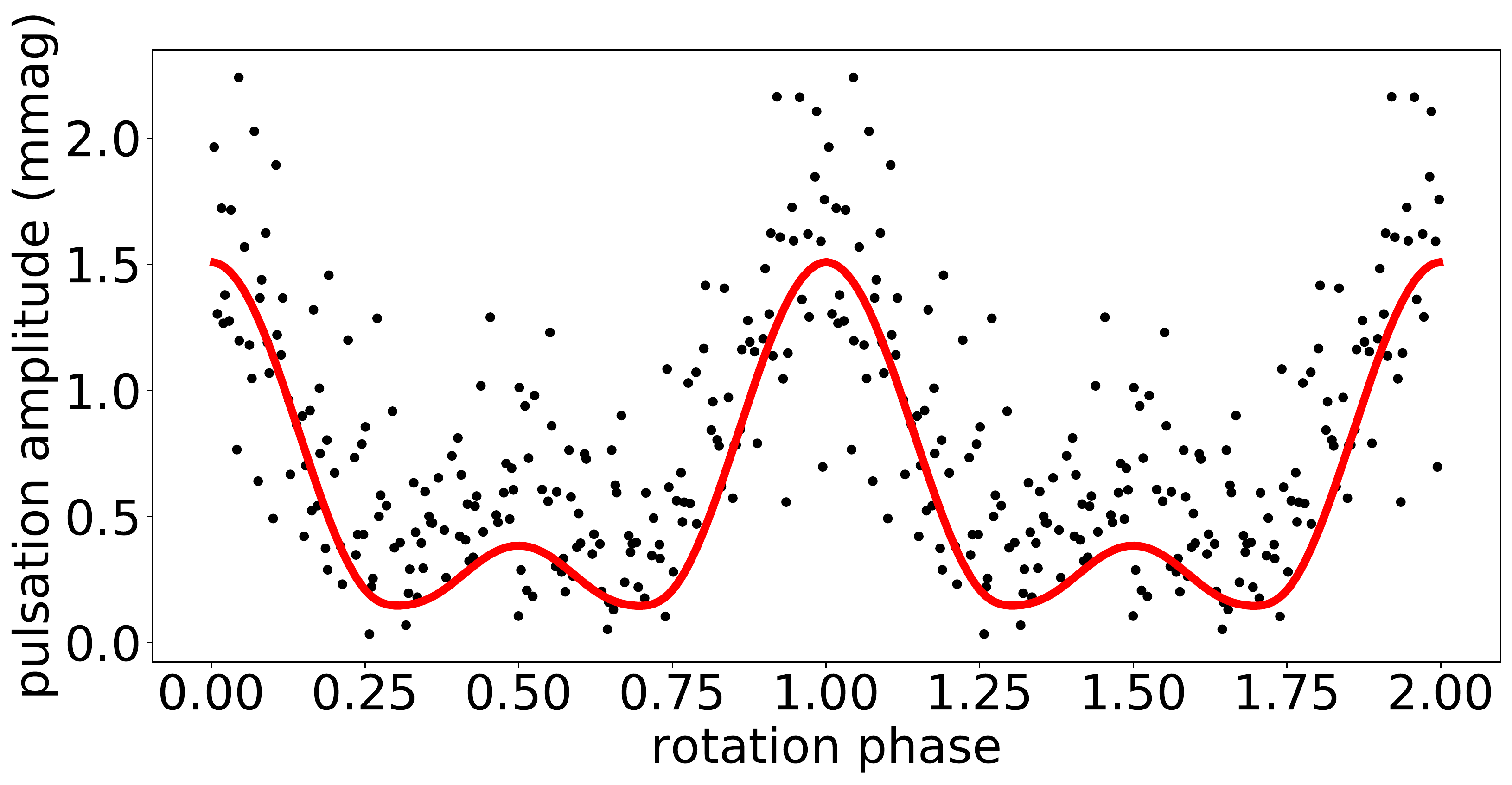}	
\includegraphics[width=1.0\linewidth]{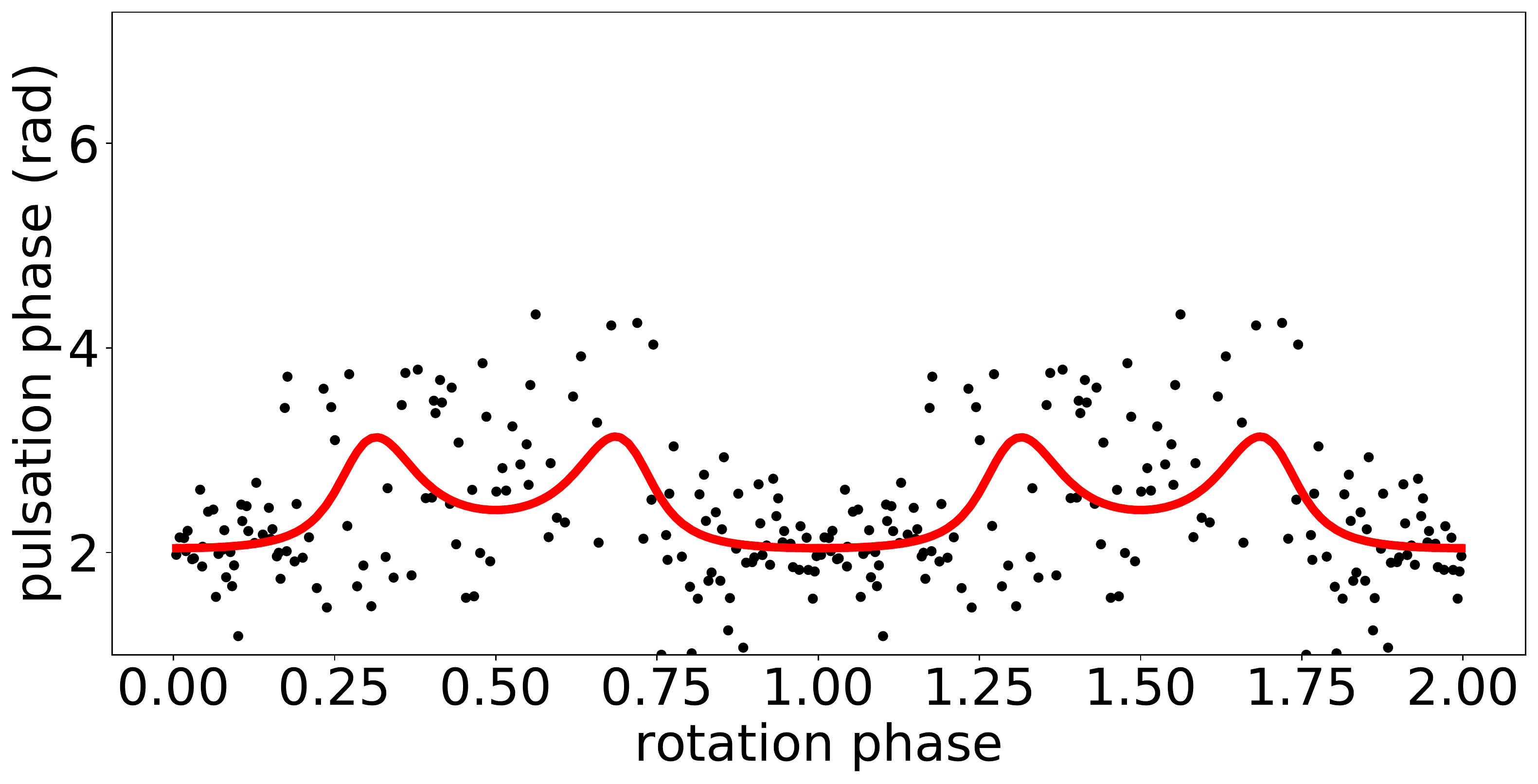}	
\caption{Same as Fig.\,\ref{fig:modulation} for KIC~10685175. The time zero-point is $t_0 = {\rm BJD}~2458711.21931$. }
\label{fig:modulation_kic}
\end{figure}

At rotation phase 0, which we chose to be the time of pulsation maximum for the quadrupole mode, we see that the spots show the secondary rotational light maximum. In contrast, for another roAp star with a quadrupole mode, KIC~10685175, the maximum of the pulsation amplitude coincides with the minimum of the rotational light \citep{2020ApJ...901...15S} (Fig.\,\ref{fig:modulation_kic}). 

The pulsation phase as a function of rotation does not show a $\pi$-rad phase reversal expected at the times of amplitude minima as would be the case for an undistorted mode, although the pulsation phase shows bumps at those times. This then argues for a distorted quadrupole mode, and also is similar to what is observed in other roAp stars with well-studied quadrupole modes \citep{2019MNRAS.489.4063H,2018MNRAS.476..601H,2018MNRAS.480.2405H,2018MNRAS.473...91H,2014MNRAS.443.2049H,1996MNRAS.281..883K,2016MNRAS.462..876H}. 

\begin{figure*}
\centering
\includegraphics[width=0.4\linewidth]{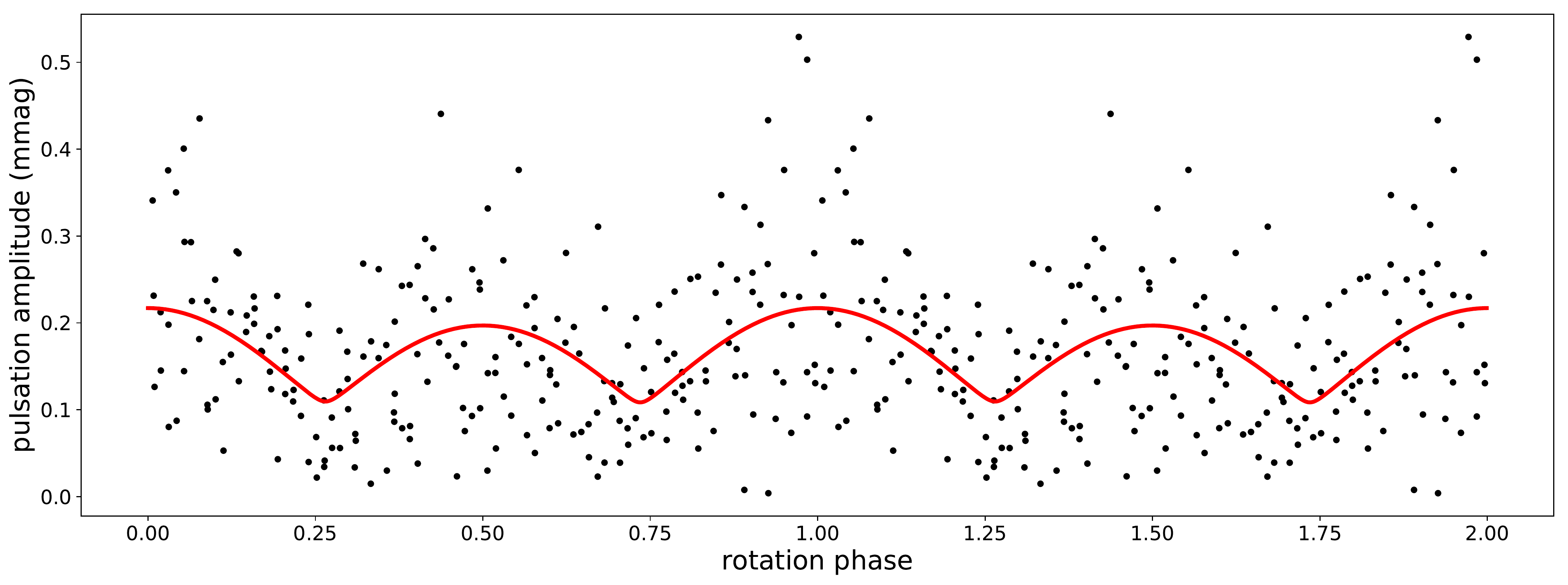}	
\includegraphics[width=0.4\linewidth]{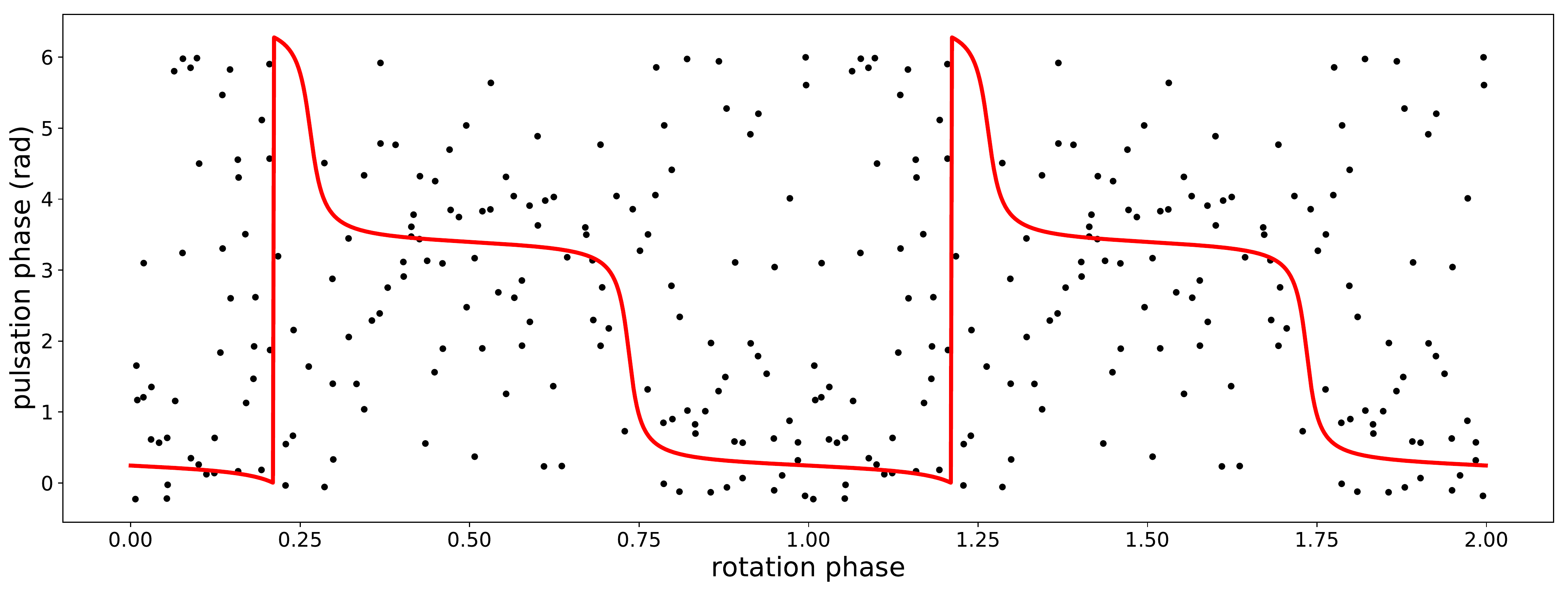}	
\includegraphics[width=0.4\linewidth]{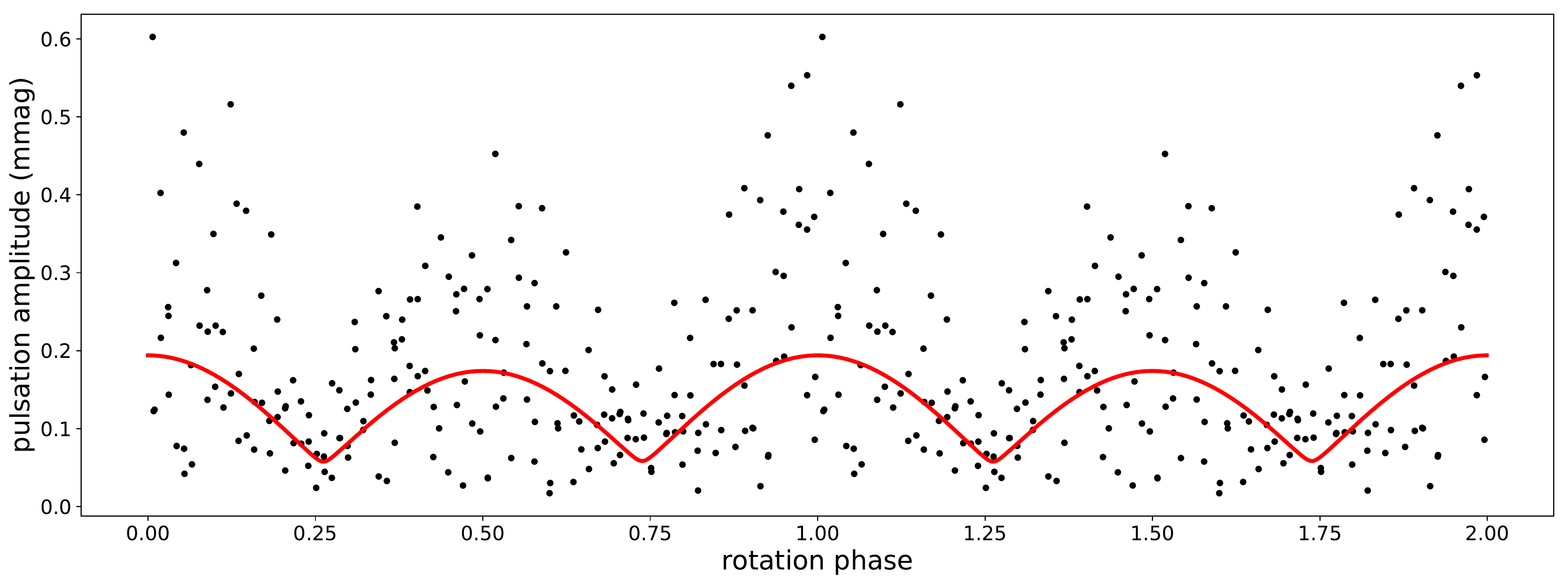}	
\includegraphics[width=0.4\linewidth]{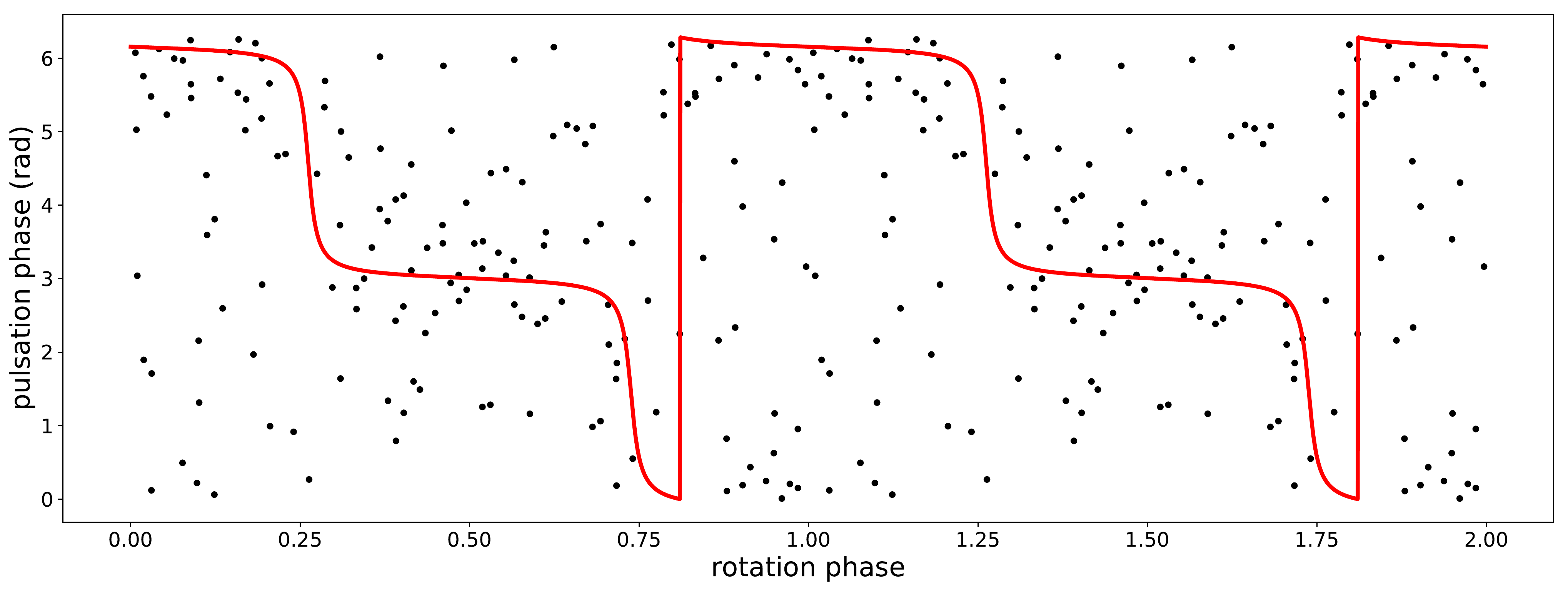}
\caption{Top panel: The pulsation amplitude (left) and phase (right) modulation of the dipole central frequency $\nu_2=230.1038$\,d$^{-1}$. Bottom panel: The pulsation amplitude (left) and phase (right) modulation of the dipole central frequency $\nu_3=235.7370$\,d$^{-1}$. Phase points with {$1\sigma$} errors greater than 1.0 rad are not plotted here. The red lines are theoretical amplitude modulation modelled following \citet{1992MNRAS.259..701K} with the components from Table~4. The time zero-point is $t_0 = {\rm BJD}~2458569.26128$. }
\label{fig:modulation_f2f3}
\end{figure*}

We also checked the pulsation amplitude and phase modulations of the two central frequencies ($\nu_2=230.1038$\,d$^{-1}$ and $\nu_3=235.7370\,$d$^{-1}$) of the dipole modes, as seen in  Fig.\,\ref{fig:modulation_f2f3}. However, because of the low amplitudes, the modulation curves are quite scattered, especially the phase modulation curve. The two dipole modes show similar behaviour: the pulsation amplitude reaches primary and secondary maximum at rotation phases 0 and 0.5, respectively, the same as the quadrupole mode. The pulsation phase variations have large errors, hence $\pi$-rad pulsation phase changes at rotation phases 0.25 and 0.75 -- typical behaviour for dipole modes -- are neither ruled out, nor supported by the plots in Fig.\,\ref{fig:modulation_f2f3}.

\section{Oblique pulsator model}
\label{ib_cal}

The oblique pulsator model describes the pulsation pattern of an oblique pulsator and only considers the surface geometry of non-radial pulsation modes. However, some spectroscopic observations (e.g. \citealt{2006AandA...446.1051K,2009MNRAS.396..325F}) and simulations (e.g. \citealt{2009ApJ...704.1218K}) have shown that properties of pulsations change rapidly with height in the stellar atmosphere and modes are substantially distorted by the magnetic field. \citet{2011MNRAS.414.2576S} and \citet{2018MNRAS.480.1676Q} have also studied this extensively theoretically. 

Recently, TESS observations of HD~6532 and HD~80316 \citep{2021arXiv210513274H} have shown that there are changes in multiplet structure comparing to the former ground-based $B$ observations. The TESS filter is broad-band white-to-red, which probes to a different depth in the stellar atmosphere than the $B$ filter. These new observations show the complexity of roAp pulsations and importance of the vertical dimension. Nevertheless, the oblique pulsator model still allows us a simple first look at the geometry of the pulsation modes.

For a normal quadrupole pulsator, the ratio of the sidelobes to the central peak can be calculated with eqns 8 and 10 from \citet{1992MNRAS.259..701K}: 
\begin{equation}
\frac{A_{+1} + A_{-1}}{A_0}=\frac{12 \sin \beta \cos \beta \sin i \cos i}{(3 \cos^2\beta -1)(3 \sin^2i-1))}
\end{equation}
\noindent and
\begin{equation}
\frac{A_{+2} + A_{-2}}{A_0}=\frac{3 \sin^2 \beta  \sin^2 i }{(3 \cos^2\beta -1)(3 \sin^2i-1))}.
\end{equation}

\noindent Dividing the two equations leads to a standard constraint for oblique pulsators with quadrupole modes:

\begin{equation}
\tan i \tan \beta =4\frac{A_{+2} + A_{-2}}{A_{+1} + A_{-1}}.
\end{equation}

We can calculate the rotation inclination $i$ and magnetic obliquity $\beta$ of a quadrupole pulsator. Although this relation applies in the case of a pure quadrupole mode, the results can provide us some information about the geometry of the mode in HD~86181 for the pure case.

The determination of $\tan i \tan \beta$ for a dipole mode is similar to that shown in eqn 3, but it is not possible to constrain $i$ and $\beta$ independently. However, for a normal quadrupole mode, eqns 1 and 2 provide two equations in two unknowns, allowing us nearly uniquely to derive values for $i$ and $\beta$. From eqns 1 and 3  we find $i = 84 \pm 3^\circ$, $\beta = 30 \pm 3^\circ$, or vice versa, for HD~86181. The uncertainties were calculated through MCMC fits.  With $i$, together with the rotation period and the estimated radius, $v\sin i = 6.5$\,km~s$^{-1}$  can be derived. Although there is no published $v\sin i$ value for comparison, this value is reasonable for a roAp star.

\begin{figure*}
\includegraphics[width=1.0\linewidth]  {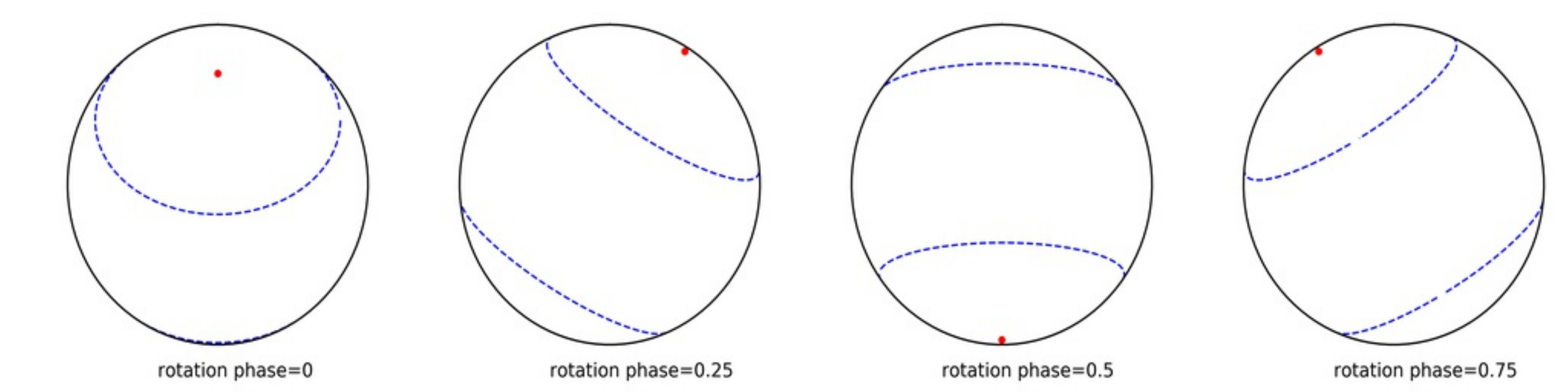}
\caption{The schematic diagram of the viewing geometry of quadrupole mode of HD~86181 through one rotation cycle. Red dots indicate the pulsation poles, and blue dashed lines indicate the surface nodes at co-latitudes $\pm 54.7^\circ$. }\label{fig:schematic}
\end{figure*}

For an axisymmetric quadrupole mode, the pulsation amplitude at the poles is twice that at the equator and in antiphase. Maximum pulsation amplitude for the angles determined above comes when $i-\beta = 54^\circ$. Since the surface nodes for an $\ell=2, m=0$ quadrupole lie at co-latitudes $\pm 54.7^\circ$, at the time of pulsation maximum the pole is inclined $i-\beta = 54^\circ$ to the line of sight, one surface node is tangent to the lower limb of the star, and the other surface node is over the the top limb. Hence we are seeing only the pulsation polar cap at that time. Half a rotation later, the pole is inclined by $i + \beta = 114^\circ$; i.e., the pole we were seeing is now on the other side of the star. The second pole has come into view, but is at poorer viewing aspect, being inclined $66^\circ$ to the line of sight. That then puts one of the surface nodes close to the line of sight, i.e. $66 - 54 = 12^\circ$. Hence much of the visible hemisphere is dominated by the equatorial region. Figure~\ref{fig:schematic} shows schematically this geometry at four rotation phases.

For a pure oblique quadrupole mode, the pulsation amplitude distribution on the surface, $A_{\theta}$, is proportional to $\frac{1}{2}(3\cos^2\theta -1)$ where $\theta$ is co-latitude, the angle to the poles. With knowledge of the rotational inclination, $i$, and magnetic obliquity, $\beta$, we can calculate an integral to obtain the pulsation amplitude at any time during a rotation cycle. Numerically, the sphere surface of the star is divided into a grid; then, with the formula, the pulsation amplitude for each cell of the grid can be calculated. With $i$, $\beta$ and the rotation angle at a given time (t), $2{\pi}{\nu}_{rot}t$, we know which grid cells can be seen by us and also the projection of each cell. Then the integrated and projected surface pulsation amplitude can be derived. The limb-darkening model for TESS \citep{2018A&A...618A..20C} is used here. The results are shown as the blue curves in Fig.\,\ref{fig:modulation}. The maximum of the integral pulsation amplitude is fixed to be the same as the one derived from the model (red line). Since the calculation just considers the pulsation as a pure quadrupole mode, the difference between blue and red line shows the contribution from the radial and dipole component. 

\section{Spherical harmonic decomposition}
\label{harmonic}

Using the technique of \citet{1992MNRAS.259..701K}, the quintuplet for HD~86181 can be decomposed into a spherical harmonic series. 
This model is also based on the oblique pulsator model. Although there are some caveats of this model, estimates of pulsation amplitudes at some special phases and pulsation amplitude ratios can be made easily. 

The decomposition was done using the frequencies, amplitudes and phases from Table~\ref{Tab:ls}. In order to interpret the two maximum pulsation amplitudes, we calculated the decomposition with the time zero point $t_0 = {\rm BJD}~2458569.26128$. The results are shown in Table~\ref{Tab:harmonic}.

\begin{table}
\centering
\caption{Results of the spherical harmonic decomposition (with the time zero point $t_0 = {\rm BJD}~2458569.26128$) of the quadrupole mode in HD~86181 for $i=84^\circ$ and $\beta = 30^\circ$. }
\begin{tabular}{|p{0.8cm}|p{0.8cm}|p{0.8cm}|p{0.8cm}|p{0.8cm}|p{0.8cm}|p{0.8cm}|}
\hline
$\ell$  & $A^{(\ell)}_{-2}$ (mmag) & $A^{(\ell)}_{-1}$ (mmag) & $A^{(\ell)}_{0}$ (mmag) & $A^{(\ell)}_{+1}$ (mmag) & $A^{(\ell)}_{+2}$ (mmag)  & $\phi$ (rad)
\\
\hline
\hline
2 & 0.093 & 0.060 & $-0.278$ & 0.056 & 0.081 & -0.322\\
1 &   & 0.015 & 0.005 & 0.014 & & 1.962\\
0 &   &   & 0.542 &   &   &-0.204\\
 \hline
\hline

\end{tabular}
\label{Tab:harmonic}
\end{table}

In recent works, we have corrected a small error in the decomposition code. The original code used to calculate the decomposition of HD~6532 \citep{1996MNRAS.281..883K} and several stars miscoded equations (8) and (10) in \citet{1992MNRAS.259..701K}. 

The decomposition components of HD~86181 show that at phase = 0, the dipole $\ell = 1$ component contributes only 0.034 mmag to the quadrupole mode -- almost nothing comparing to the strong radial contribution, which means that the polar amplitude is increased and the equatorial amplitude is reduced compared to a pure quadrupole mode. 
These results verify the assumption that the pulsation amplitude maximum comes from the poles, with the secondary maximum from the equator. 

As an example, we estimate the pulsation amplitude maximum at phase 0. 
According to the eqns (20), (21) and (22) in \citet{1992MNRAS.259..701K}, at phase 0, the pulsation amplitude is 

$A=\sqrt{(\sum\limits_{\ell=0}^{2}\sum\limits_{m=-\ell}^{\ell}A^{\ell}_m\cos{\phi^{\ell}})^2+(\sum\limits_{\ell=0}^{2}\sum\limits_{m=-\ell}^{\ell}A^{\ell}_m\sin{\phi^{\ell}})^2}$. 

\noindent{  The amplitude of the dipole mode ($A^{\ell=1}_m$) is negligible, and the quadrupole and radial components have similar phases, meaning $\phi^{\ell=2}$ and $\phi^{\ell=0}$ can be considered as the same, so they can add at the time of amplitude maximum. Therefore, the pulsation amplitude is $A=0.542+0.093+0.060-0.278+0.056+0.081=0.559$\,mmag, which fits the pulsation phase plot well. Of course, the decomposition technique was designed to fit the data, so it is not a surprise that it does. This discussion is to give a mental picture of why this is so. More precisely, a fit of all three spherical harmonic components taking into account that the exact phases seen in Table~4 gives the fit shown in Fig.\,\ref{fig:modulation} as the red curves.}

In addition to the quintuplet for HD~86181, there are two doublets. With the $i$ and $\beta$ in section~\ref{saio}, we derive $\tan i \tan \beta = 4.76$. For dipoles that gives

\begin{equation}
\frac{A_{+1} + A_{-1}}{A_0} = 4.76.
\end{equation} 

\noindent We therefore expect to see triplets with very small central components at $\nu_2$ and $\nu_3$, with amplitudes only about 0.03\,mmag, which is at the detection limit for these data. This supports the identification of $\nu_2$ and $\nu_3$ as dipole modes, and it is therefore no surprise that we see doublets separated by twice the rotation frequency.

\section{The large separation and acoustic cut-off frequency}
\label{sep}

The large separation, $\Delta \nu$, is the separation in frequency of modes of the same degree and consecutive radial orders, and is proportional to the square-root of the mean density of the star, i.e., $\Delta \nu \propto \sqrt{\rho}$  (e.g. \citealt{1985AandA...143..206G}). This relation was developed for the frequencies of high-order, acoustic, adiabatic, non-radial oscillations \citep{1980ApJS...43..469T,1990ApJ...358..313T}. Since the roAp pulsations are in the asymptotic regime, they are also applicable here. If the pulsation modes, or at least relative radial orders are identified, the large separation can, in principle, be determined. 

To calculate the large separation, stellar radius and mass are required. We estimate the radius of HD~86181 from $L=4{\pi}{\sigma}R^2T_{\rm eff}^4$, and its mass from $M/{\rm M}_{\odot}=(L/{\rm L}_{\odot})^{1/4}$ (derived from stellar homology relations; see, e.g., \citealt{1924Natur.113..786E}) with the luminosity in Table~\ref{Tab:param}. We find $R=1.65$\,R$_{\odot}$, $M = 1.72$\,M$_{\odot}$, and $\log g = 4.19$\,(cgs) for HD~86181. Although the mass is obtained from a rough scaling relation, it is fine for estimating the large separation and the cut-off frequency in this section.


With the knowledge that the doublets we see are the result of dipole modes with undetected central peaks, we are able to derive the mode frequencies to be $\nu_2 = 230.103$\,d$^{-1}$ and $\nu_3 = 235.737$\,d$^{-1}$ by taking the average of the two sets of sidelobes. That then gives the mode frequency separations to be $\nu_1 - \nu_2 = 2.668$\,d$^{-1}$ = 30.87\,$\umu$Hz, and $\nu_3 - \nu_1 = 2.967$\,d$^{-1}$ = 34.32\,$\umu$Hz. Using the {radius, mass and $\log$g estimated above} and the value of the solar large frequency separation $\Delta \nu_{\odot} =134.88\,\umu$Hz \citep{2011ApJ...743..143H}, through $\Delta \nu \propto \sqrt{\frac{g}{R}}$, we estimate $\Delta \nu /2 = 38.2$\,$\umu$Hz, which is consistent with the observations.

In roAp stars part of the pulsation mode energy can be refracted back into the star by the influence of the magnetic field, even when the frequency of the mode is above the acoustic cut-off frequency, $\nu_{ac}$ \citep{sousa08,quitral18}. Therefore, there is no reason to assume that very high frequency modes will not be observed in these pulsators. Nevertheless, theory predicts that the excitation by the opacity mechanism takes place in a frequency range that is close to, but does not exceed the cut-off frequency and, thus, that an alternative excitation mechanism would be required to excite modes of yet higher frequencies \citep{2013MNRAS.436.1639C}. It is therefore of interest to estimate the cut-off frequency in HD~86181 based on the star's global properties. Using the mass, radius and the effective temperature in solar values in Table~1, and the scaling relation $\nu_{ac} \propto g/\sqrt{T_{\rm eff}}$ \citep{1991ApJ...368..599B} with $\nu_{ac,\odot}=5.55$\,mHz \citep{1992AandA...266..532F}, we find that in HD~86181 $\nu_{ac} \approx3.03$\,mHz, which is slightly larger than the observed mode frequencies, around 2.73\,mHz.

\section{Modelling oblique quadrupole pulsations distorted by dipole magnetic fields}
\label{saio}

In this section, we present comparisons of the observed amplitude and phase modulations of HD~86181 with a quadrupole pulsation calculated by the method of \citet{2005MNRAS.360.1022S} including the effect of a dipole magnetic field. We assume that the pulsations in roAp stars are axisymmetric with the pulsation axis aligned with the axis of the dipole magnetic field. The strength of the field is denoted by $B_{\rm p}$, the magnetic field strength at the poles.  

\begin{figure}
\includegraphics[width=0.5\textwidth]{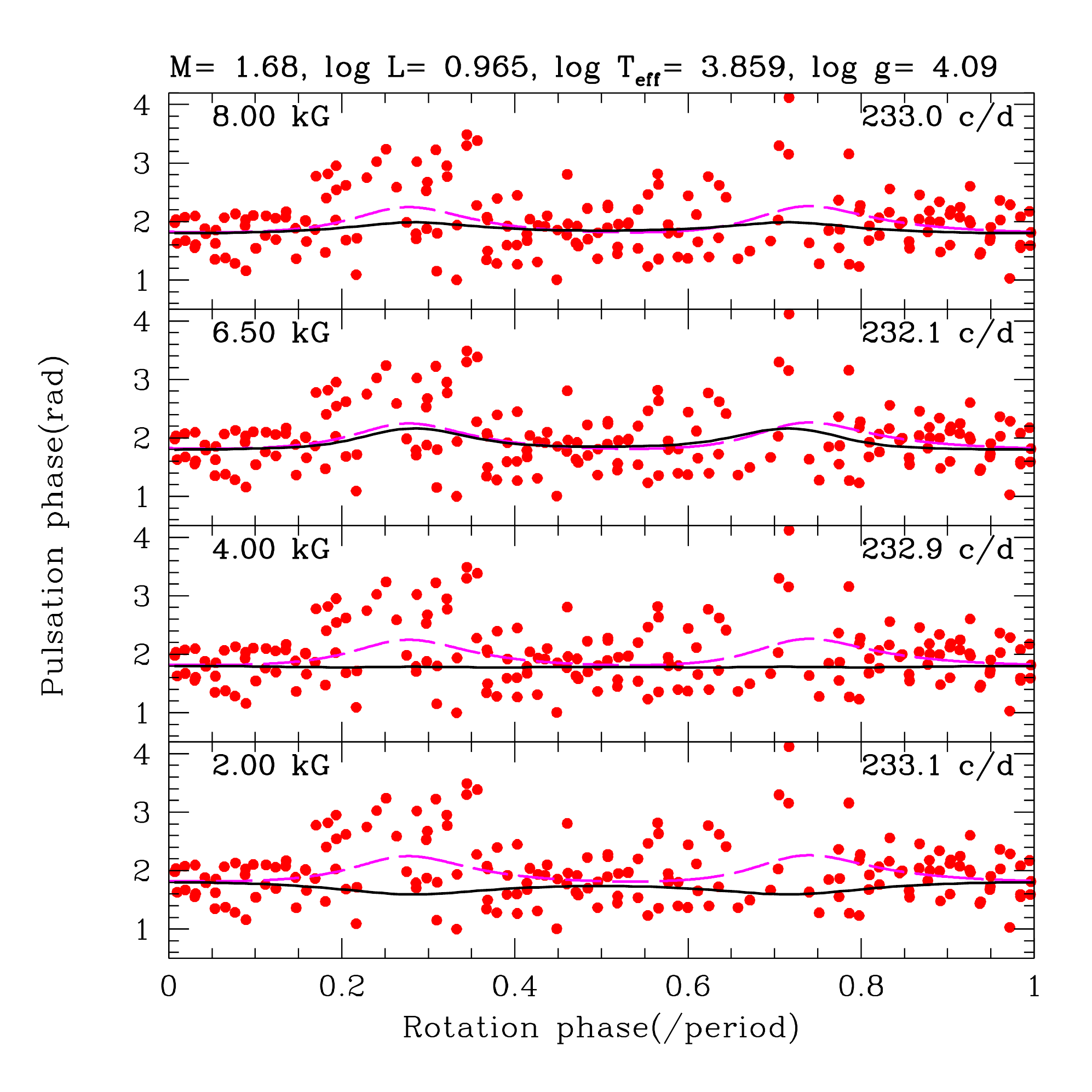}
\caption{Phase modulations (solid black lines) obtained by assuming various strengths of magnetic fields for the quadrupole mode in the same model shown in Fig.~\ref{fig:model_yuki}. Red dots are observed phase modulations of HD~86181, while dashed magenta lines are the same as the one in the bottom panel of Fig.~\ref{fig:model_yuki}, which are obtained from the oblique pulsator model of \citet{1992MNRAS.259..701K} (red lines in Fig.~\ref{fig:modulation}). For all cases, $(\beta,i)=(40^\circ,80^\circ)$ are adopted, for which the theoretical amplitude modulations are consistent with that of HD~86181, while B$_p$ in the range 6.5 - 8\,kG (e.g. 7\,kG; Fig.~\ref{fig:model_yuki}) gives phase modulation comparable with the observed one (see also Fig.~\ref{fig:model_yuki}).}
\label{fig:phase}
\end{figure}

In the presence of a magnetic field, the pulsation frequency is modified only slightly (see Fig.~\ref{fig:phase}), while the eigenfunction is distorted considerably because the magnetic effect generates $\ell = 0, 4, 6,  \ldots$ components of spherical harmonics in addition to the main $\ell=2$ component. (We have included twelve components; i.e, up to $\ell = 22$.) The eigenfunction gives pulsation amplitude and phase on each point on the surface as a function of the angle from the magnetic (or pulsation) axis. The amplitude/phase distribution can be converted to observational amplitude/phase modulation as a function of rotation phase (see \citealt{2004MNRAS.350..485S} for details) for a set of $(\beta, i)$.  The method of comparison is also discussed in \citet{2020ApJ...901...15S}.

\begin{figure}
\includegraphics[width=0.5\textwidth]{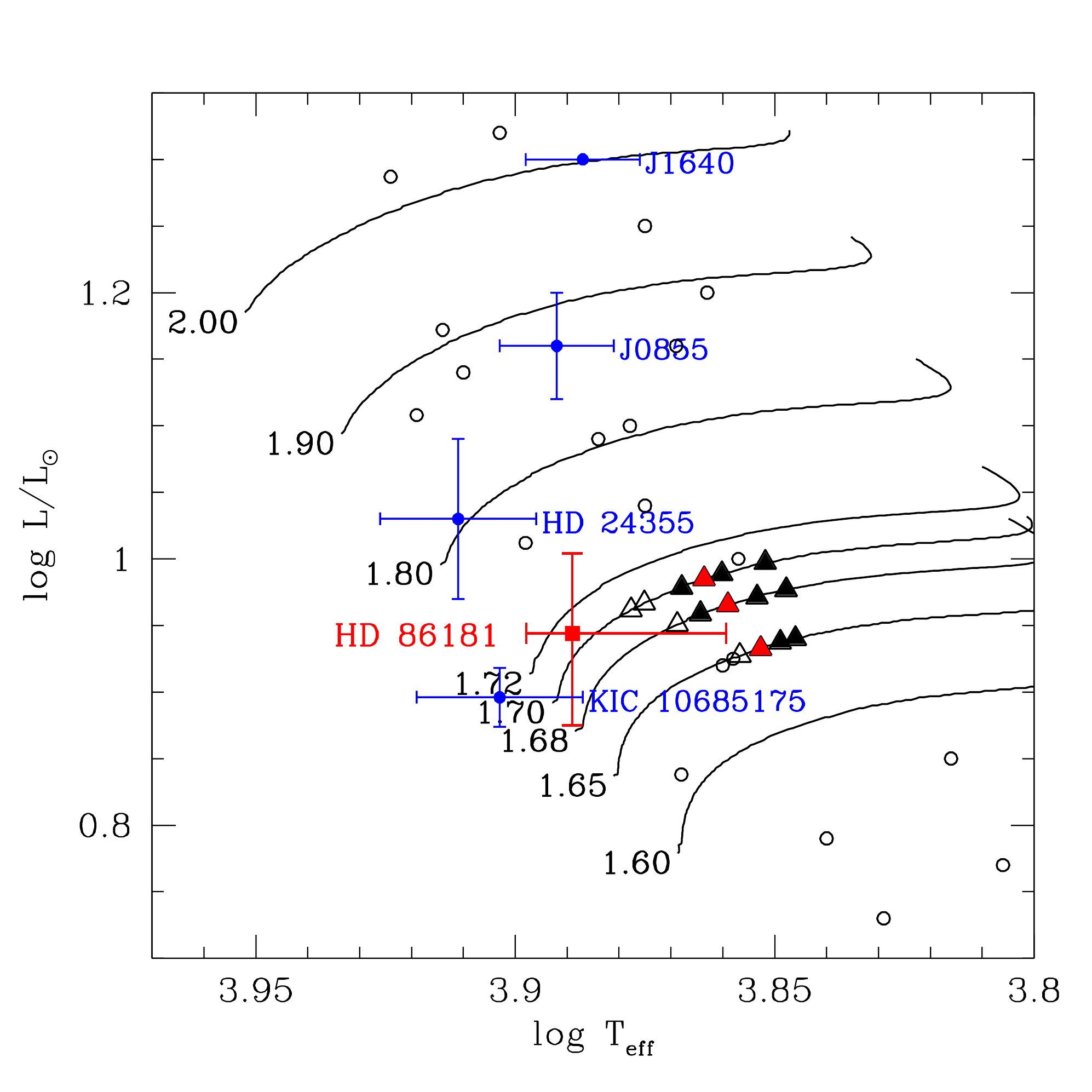}
\caption{Loci of roAp stars on the HR diagram with some evolutionary tracks with initial composition of $(X,Z)=(0.70,0.02)$. The number along the ZAMS of each track indicates the stellar mass in solar units. HD~86181 is shown in red and other distorted quadrupole pulsators are shown in blue for comparison. (J1940 is not shown because its location is very close to J1640.) Triangles on 1.70, 1.68 and 1.65\,M$_\odot$ tracks indicate the loci of models for which pulsation amplitude and phase modulations are calculated; filled (open) triangles indicate models whose phase modulations can (cannot) be fitted with the HD~86181 phase modulation. Red filled triangles indicate models which have large frequency spacings similar to the observed one.  Parameters of roAp stars other than HD~86181 are adopted from \citet{2018MNRAS.476..601H}. }
\label{fig:hrd}
\end{figure}

According to the estimated luminosity range, we selected some models on the 1.65, 1.68, and 1.70\,M$_\odot$ evolutionary tracks as indicated by triangles in the HR diagram of Fig.\,\ref{fig:hrd}, in which the initial composition $(X,Z)=(0.70,0.02)$ is adopted, while the helium abundance is assumed to be depleted to 0.01 (mass fraction) in the layers above the second helium ionisation zone (polar model in \citealt{2001MNRAS.323..362B}). For a stellar model, we find, firstly without including a magnetic field, a quadrupole mode having a pulsation frequency close to $\nu_1 =232.77\,{\rm d}^{-1}$. Then, we re-calculate the quadrupole mode by taking into account the effect of an assumed dipole magnetic field of $B_{\rm p}$.

For each case, an appropriate set of $(\beta,i)$ is determined by fitting the amplitude modulation of HD~86181. Then, the phase modulation is compared with the observations.  Generally, for most assumed values of  $B_{\rm p}$, the obliquity and inclination angles $(\beta,i)$ can be determined easily by fitting the predicted amplitude modulation with the observations, while the theoretical phase modulation tends to be very small except for  a certain range of  $B_{\rm p}$. Fig.~\ref{fig:phase} shows how theoretical phase modulations change with changing $B_{\rm p}$ for a 1.68\,M$_{\odot}$ model. In this model, $6.5 \lesssim B_{\rm p}$/kG $\lesssim 8$ gives phase modulations that are comparable with the observed ones. 
The required $B_p$ tends to be smaller in more massive models because the mean density of the envelope is smaller in more massive stars.

Filled triangles in Fig.~\ref{fig:hrd} indicate the loci of models whose amplitude and phase modulations agree with the observed ones of HD~86181; agreements occur if  $B_{\rm p} \sim 9.0 - 6.0$~kG is assumed depending on the assumed stellar mass of 1.65, 1.68, 1.70-M$_\odot$. Among them, the three red triangles denote the models whose large frequency separations agree with that of HD~86181. We have chosen the 1.68-M$_\odot$ model as the best model because the luminosity agrees with our derived value better than the luminosity of the 1.70-M$_\odot$ model does.  However, $\log T_{\rm eff}=3.859$  ($T_{\rm eff}=7230$\,K) of the most approriate fit model is somewhat lower than 7750\,K listed in Table~\ref{Tab:param}. This $T_{\rm eff}$ value is closer to $T_{\rm eff}= 7320$ K obtained by \citet{2012MNRAS.427..343M} from a comparison of the SED with model atmospheres, and to $T_{\rm eff}=7205$\,K obtained by \citet{2006AandA...450..735M} from 2MASS photometry.

\begin{figure}
\includegraphics[width=0.5\textwidth]{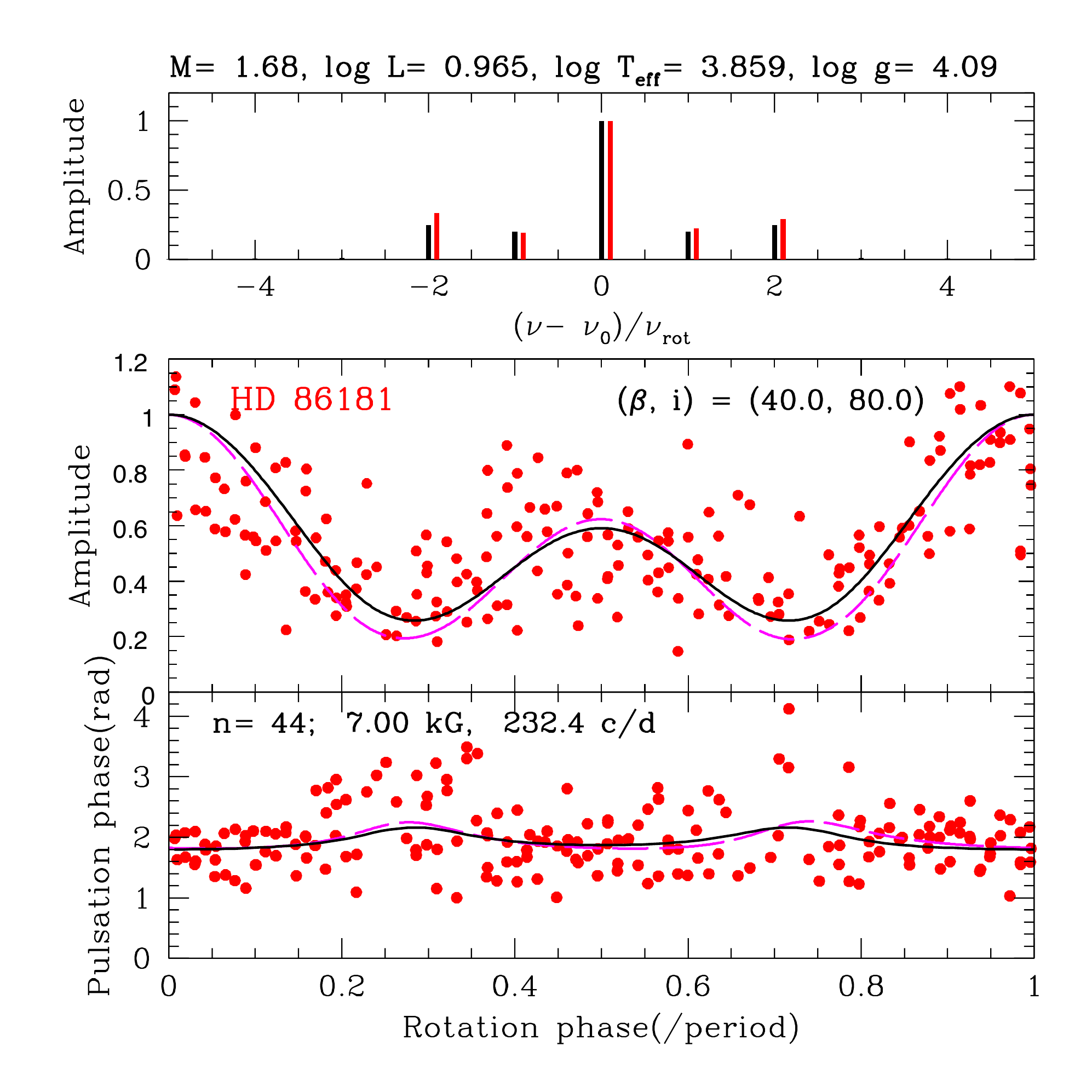}
\caption{The amplitude spectrum of rotational sidelobes (top panel) and amplitude (middle panel)/phase (bottom panel) modulations of the quadrupole pulsation mode of HD~86181 are shown by red lines or dots. Dashed magenta lines (middle and bottom panels) are obtained from the oblique pulsator model of \citet{1992MNRAS.259..701K} (red line in Fig.~\ref{fig:modulation}). Black lines show the results of a best model of 1.68\,M$_\odot$ with $B_{\rm p}= 7$~kG, for which parameters are shown on the top of the diagram.}
\label{fig:model_yuki}
\end{figure}

Fig.~\ref{fig:model_yuki} compares amplitudes of the rotational sidelobes (top), amplitude (middle) and phase (bottom) modulations between the best model with $B_{\rm p}= 7.0$\,kG and HD~86181.By fitting the amplitude modulation, we find $(\beta, i)$ = ($40^\circ$,$80^\circ$) each with an uncertainly of 5$^\circ$. The $(\beta, i)$ given by the magnetically distorted model are only slightly different from the pure quadrupole pulsator model: $i$ given by the distorted model is consistent with the pure quadrupole pulsator model within the 1$\sigma$, while $\beta$ is consistent within 2$\sigma$. The range of the phase modulation of the quadrupole model is small, which can be attributed to contributions from $\ell=$4, 6, 8, .....

The dipole mode frequencies just above and below the quadrupole mode of the best fitting model are 235.51 and 229.92~d$^{-1}$, respectively, at $B_{\rm p}= 7.0$~kG, which yield a large frequency spacing of  {5.59~d$^{-1}$ (or 64.7\,$\umu$Hz)},  which agrees with the observed large frequency spacing, $\nu_3-\nu_2=5.63$~d$^{-1}$ (or 65.2\,$\umu$Hz).\footnote{The frequency spacing of this model at $B_{\rm p}= 0$ is 5.39~d$^{-1}$ (or 62.4\,$\umu$Hz).}  

For HD~42659, another roAp star pulsating in a distorted mode \citep{2019MNRAS.489.4063H}, the distorted model predicted the polar magnetic field strength to be 0.8\,kG by assuming that star pulsates in a quadrupole mode. That result was consistent with the measured mean longitudinal magnetic field, $\langle B_l \rangle = 0.4$\,kG  \citep{2006AandA...450..763K,2006AN....327..289H}. However, the polar magnetic field strength predicted by our model for HD~86181, $B_p =  7.0$\,kG, is significantly larger than the measured mean longitudinal magnetic field, $\langle B_l \rangle = 0.54$\,kG \citep{2015AandA...583A.115B}.  
The cause of the difference is not clear. It could be a depth effect; i.e., the magnetic field required in our model refers to the strength in the hydrogen-rich envelope, while the measured magnetic field corresponds to the strength in the outermost superficial layers. Also, there are some aspects that are not considered in the model, such as the effects of surface spots. 


\section{Driving of pulsations}
\label{cunha}

The driving of pulsations in roAp stars is still a matter of debate. Non-adiabatic pulsation calculations, assuming that envelope convection is suppressed by the magnetic field at least in some angular region around the magnetic pole, have been reasonably successful in explaining the driving of most oscillations observed in roAp stars through the opacity mechanism acting on the hydrogen ionization region \citep{2001MNRAS.323..362B,cunha02}. The same model also predicts that very high frequencies may be excited by the turbulent pressure mechanism, a fact that has been suggested to explain the pulsation frequencies observed in the roAp star $\alpha$~Cir \citep{2013MNRAS.436.1639C}. In this section we adopt the models discussed in these earlier works to perform theoretical non-adiabatic pulsation calculations for HD~86181. 

The analysis follows closely that presented by \cite{2013MNRAS.436.1639C}. In short, the equilibrium model is derived from the matching of two spherically symmetric models, one with envelope convection suppressed (the polar model) and the other with convection treated according to a non-local mixing length prescription \citep{spiegel63,gough77a} (the equatorial model). It takes as input the stellar mass, luminosity, effective temperature, chemical composition (hydrogen, $X$, and helium, $Y$, mass fractions) and the parameters associated with convection. The atmosphere is described by a $T-\tau$ relation, which can be chosen amongst different options, with the minimum optical depth, $\tau_{\rm min}$, being an additional input parameter. Finally, helium settling can also be considered both in the polar and in the equatorial regions, following a parametrized description with the surface helium abundance in each region being additional input parameters. 

The stability analysis is performed in each region separately and can consider two different options for the surface boundary condition applied at the minimum optical depth, namely, one that guarantees a full reflection of the mode and one that allows waves with frequencies above the acoustic cut-off frequency to propagate. In the equatorial model, the final non-adiabatic solutions are computed using a non-local, time-dependent mixing-length treatment of convection \citep{gough77b, balmforth92}. The results from the non-adiabatic analysis in each region can then be combined to derive the growth rates of modes in the model where convection is assumed to be suppressed only in some angular region around the magnetic pole (the composite model). Further details on the models can be found in \cite{2001MNRAS.323..362B} and references therein.

For each set of ($M$,$L$,$T_{\rm eff}$), four different physics configurations were considered by varying different input parameters identified in previous works as having significant impact on the stability results, namely: the minimum optical depth, the outer boundary condition, and the amount of surface helium. Table~\ref{tab:models} summarizes the options in each case. Other parameters and physics not mentioned here were fixed following the options adopted in \cite{2001MNRAS.323..362B}.

\begin{table*}
  \centering
  \caption{Modelling parameters for the cases illustrated in Fig.\,\ref{fig:model}, all computed with $M=1.72\,{\rm M}_{\odot}$, $L=8.69\,{\rm L}_{\odot}$, $Y=0.278$, $X=0.705$.}
  \label{tab:models}
  \begin{tabular}{cccccc}
    \hline
    Model & Polar $Y_{\rm surf}$ & Equatorial $Y_{\rm surf}$ & $\tau_{\rm min}$  &Boundary & symbols  \\
                  &          &      && condition & in Fig.\,\ref{fig:model}  \\
    \hline
    A & 0.01 & 0.278 & $3.5\times10^{-5}$   &  Reflective & circles\\
    B  & 0.01  & 0.278 & $3.5\times10^{-4}$  &  Reflective & squares \\
    C & 0.01   & 0.278 & $3.5\times10^{-5}$  & Transmissive & upward triangles\\
    D &   0.1  & 0.1 & $3.5\times10^{-5}$  &  Reflective & rightward triangles\\

  \hline
  \end{tabular}
\end{table*}

\begin{figure}
\centering
\includegraphics[width=1.0\linewidth]{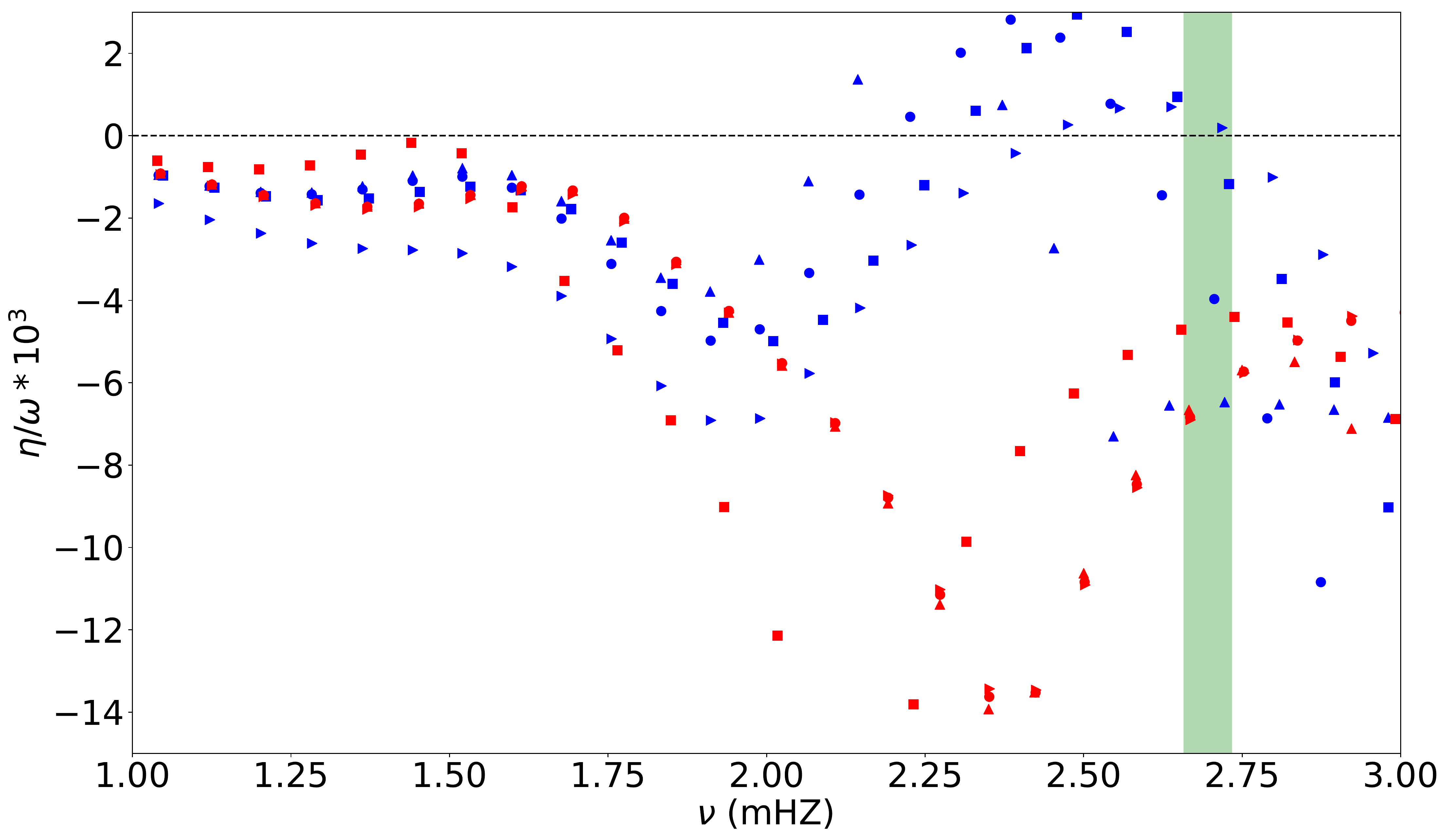}	
\caption{Normalized growth rates for polar (blue) and equatorial (red symbols) regions as a function of the cyclic frequency $\nu=\omega/2\pi$. Excited modes have positive growth rates. Different shape symbols represent different modelling parameters we have used; circles are for model A in Table~\ref{tab:models}, while squares, upward triangles and rightward triangles for model B, C and D, respectively. Zero growth rate is indicated by the horizontal dashed line and the green shadowed region marks the range of observed frequencies.}
\label{fig:model}
\end{figure}

Fig.\,\ref{fig:model} shows an example of the results from the stability analysis in blue and red, for polar and equatorial models, respectively, adopting the effective temperature and luminosity in Table~\ref{Tab:param}. Here we plot the relative growth rates $\eta/\omega$ as a function of the cyclic pulsation frequency $\nu$, where $\eta$ and $\omega$ are the imaginary and real parts of the angular eigenfrequency, respectively, and a positive growth rate indicates the mode is intrinsically unstable, thus excited. From the red symbols in the figure we can see that all modes are stable in the equatorial model, independently of the physics configuration adopted. In the polar models (blue symbols), a few modes have positive growth rates at frequencies from $\sim$ 2.1~mHz up to $\sim$~2.7~mHz, depending on the physics considered. The range of excited frequencies scales approximately with the square root of the mean density \citep{cunha02,2013MNRAS.436.1639C}. Given the uncertainty on the radius of the star, one can thus confidently conclude that the region where the oscillation frequencies are observed is within the range where the polar models predict instability. Despite this, the growth rates on these polar models are one order of magnitude smaller than the growth rates of the corresponding modes in the equatorial model (in absolute value). This means that envelope convection needs to be almost fully suppressed in order for these modes to be unstable in the composite model \citep[cf. figure 4 of ][]{2001MNRAS.323..362B} and, thus, explain the observations.

\section{Discussion and conclusions}

We analysed HD~86181 with TESS data, and confirm it as a roAp star. The rotation frequency is derived to be $\nu_{\rm rot} = 0.48765 \pm 0.00003$\,d$^{-1}$ ($P_{\rm rot} = 2.0507 \pm 0.0001$\,d). The pulsation frequency spectrum is rich, consisting of one doublet, one quintuplet and another doublet. The central frequency of the quintuplet is 232.7701\,d$^{-1}$ (2.694\,mHz). The two doublets are very likely to be sidelobes of two triplets, the amplitudes of whose central frequencies are too small to be observed. With this interpretation, we calculate the two central frequencies of the triplets to be 230.1028\,d$^{-1}$ (2.663\,mHz) and 235.7361\,d$^{-1}$ (2.728\,mHz). 

Pulsation amplitude and phase modulation were calculated as a function of rotation phase and shown to be modulated. Two maxima can be seen in the rotational light curve, which indicates we see two primary spots in the TESS pass-band, but the spot geometry is complex and further work is needed to construct the chemical and magnetic map of this star. 

We calculated the rotation inclination, $i$, and magnetic obliquity, $\beta$, for HD~86181, which provided detailed information of the geometry and we used those values with a spherical harmonic decomposition to better understand the pulsation geometry and the distortion from a pure quadrupole mode. 

Models considering the dipole magnetic field distortion were calculated and compared with the observed amplitude and phase modulation. The best fit model gives $B_{\rm p}= 7.0$~kG and $(\beta, i) = (40^\circ,80^\circ)$. The $(\beta, i)$ given by magnetic distortion model are only slightly different from the pure quadrupole pulsator model with the relevant differences of (25~per~cent, 5~per~cent) for $(\beta, i)$, respectively. Also, the difference from the phase modulation of the quadrupole model is small, which can be attributed to higher degree components, $\ell=4, 6, 8, ...$. The pulsation frequency and the large frequency spacing given by this model are comparable with the observation.

To explain the driving mechanism of this star, two non-adiabatic models were constructed for HD\,86181, one with envelope convection suppressed (the polar model) and another considering convection (the equatorial model). We find that polar model predicted the excitation of modes in the observed range.

The rich pulsation frequency spectrum let us study the large frequency separation, $\Delta \nu$. The $\Delta \nu$ derived from $g$ and $R$ is consistent with the observed value. The acoustic cut-off frequency, $\nu_{ac}$, of this star is larger than the observed mode frequencies.

\bibliography{26ver_hd86181}{}

\section*{acknowledgements}
This work was funded by the National Key R $\&$ D Program of China under grant No.2019YFA0405504 and the National Natural Science Foundation of China (NSFC) under grants No. 11973001, No. 11833002, No. 12090040 and No. 12090042. This work includes data collected by the TESS mission. Funding for the TESS mission is provided by the NASA Explorer Program. M. S. Cunha is supported by national funds through FCT in the form of a work contract and through the research grants UIDB/04434/2020, UIDP/04434/2020 and PTDC/FIS-AST/30389/2017, and by FEDER - Fundo Europeu de Desenvolvimento Regional through COMPETE2020 - Programa Operacional Competitividade e Internacionalização (grant: POCI-01-0145-FEDER-030389). Daniel L. Holdsworth acknowledges financial support from the Science and Technology Facilities Council (STFC) via grant ST/M000877/1. Gerald Handler gratefully acknowledges funding through NCN grant 2015/18/A/ST9/00578. We thank the anonymous referee for a thorough, knowledgeable review that improved this paper.

\section*{Data Availability}
The data underlying this article will be shared on reasonable request to the corresponding author.

\end{document}